\documentclass[useAMS,usenatbib,usegraphicx]{mn2e}
\usepackage{times}
\usepackage{lscape}
\usepackage{amssymb}
\usepackage{color}

\def\kms{$\rm km\, s^{-1}$}
\def\cm3{$\rm cm^{-3}$}

\def\n0{$\rm n_{0}$}
\def\B0{$\rm B_{0}$}

\def\erg{$\rm erg\, cm^{-2}\, s^{-1}$}

\def\mc{$\mu$m}

\def\L12{L$_{12\mu m}$~}
\def\F12{F$_{12\mu m}$~}

\def\fe2{[Fe\,{\sc ii}]}
\def\s3{[S{\sc iii}]}
\def\h2{H$_{2}$}
\def\w{$W_{\lambda}$}
\def\F{$F_{\lambda}$}

\def\pp{$\pm$}

\def\ergh{$\rm erg\, cm^{-2}\, s^{-1}\,Hz^{-1}$}
\def\dg{$\ddagger$}

\title[Probing the NIR stellar population of Seyfert galaxies]{Probing the near infrared stellar population of Seyfert galaxies}

\author[Riffel et al.]{R. Riffel$^{1}$\thanks{E-mail:
riffel@ufrgs.br}, M. G. Pastoriza$^{1}$, 
A. Rodr\'{\i}guez-Ardila$^2$\thanks{Visiting Astronomer at the Infrared Telescope Facility, which is operated by the University of Hawaii
under Cooperative Agreement no. NCC 5-538 with the National Aeronautics and Space Administration, Office of Space Science, Planetary Astronomy Program.}
and C. Bonatto$^1$
\\$^{1}$Departamento de Astronomia, Universidade Federal do Rio Grande do Sul. Av. Bento Gon\c calves 9500, Porto Alegre, RS, Brazil.
\\$^2$ Laborat\'{o}rio Nacional de Astrof\'{i}sica/MCT - Rua dos Estados Unidos 154, Bairro das Nac\~oes, Itajub\'a, MG, Brazil.}

\begin{document}

\date{}
\pagerange{\pageref{firstpage}--\pageref{lastpage}} \pubyear{2008}

\maketitle

\label{firstpage}

\begin{abstract}

We employ IRTF SpeX NIR (0.8\mc\---2.4\mc) spectra to investigate the stellar population
(SP), active galactic nuclei (AGN) featureless continuum ($FC$) and hot dust properties
in 9 Sy\,1 and 15 Sy\,2 galaxies. Both the {\sc starlight} code and the hot dust as an
additional base element were used for the first time in this spectral range. We found
evidence of correlation among the equivalent widths (\w)  Si\,I\,1.59\mc\ 
$\times$ Mg\,I\,1.58\mc, equally for both kinds of activity. 
Part of the $W\rm_{Na\,I\,2.21\mu m}$ and $W\rm_{CO\,2.3\mu m}$ strengths may be
related to galaxy inclination. Our synthesis shows significant differences between Sy\,1 and Sy\,2
galaxies: the hot dust component is required to fit the $K$-band spectra of $\sim$90\%
of the Sy\,1 galaxies, and only of $\sim$25\% of the Sy\,2; about 50\% of the Sy\,2 galaxies
require a $FC$ component contribution $\gtrsim$20\%, while this fraction increases
to about 60\% in the Sy\,1; also, in about 50\% of the Sy2, the combined FC and young components
contribute with more than 20\%, while this occurs in 90\% of the Sy1, suggesting recent
star formation in the central region. The central few hundred parsecs of our galaxy sample
contain a substantial fraction of intermediate-age SPs with a mean metallicity near solar. Our SP 
synthesis confirms that the 1.1\mc\ CN band can be used as a tracer of intermediate-age
SPs. The simultaneous fitting of SP, $FC$ and hot dust components increased in $\sim150\%$
the number of AGNs with hot dust detected and the mass estimated. The NIR emerges as an excellent
window to study the stellar population of Sy\,1 galaxies, as opposed to the usually heavily
attenuated optical range. Our approach opens a new way to investigate and quantify the individual
contribution of the three most important NIR continuum components observed in AGNs.

\end{abstract}

\section{Introduction}

A key issue in modern astrophysics is to understand the origin of 
the energy source that powers the continuum and line-emitting gas in 
active galactic nuclei (AGN).  The current paradigm proposes that accretion of material onto 
a supermassive black-hole located at the centre of the galaxy is the mechanism responsible for the 
observables associated with the AGN phenomena. In addition to this central engine,   
observational evidence over the past decade has shown 
that massive star forming regions are commonly detected in the central region of galaxies 
harbouring an AGN \citep[e.g.][]{tok91,msm94,id00,thaisa00,iman02,ara03,riffel07,oli08}. In this 
scenario, black-holes and starburst clusters coexist in the nuclear region of galaxies.
Currently, there is ample evidence indicating that both the active nucleus and starburst might 
be related to gas inflow, probably triggered by an axis-asymmetry perturbation 
like bars, mergers or tidal interactions \citep{sbf89,sbf90,maiolino97,ksp00,fathi06,rogemar08}, 
providing support to the so-called AGN-starburst connection
\citep[][and references therein]{ns88,terlevich90,heckman97,cid97,gd98,veilleux00,fmr00,hec04,rogemar08}.

Another line of thought, however, claims that this AGN-starburst connection could be incidental,
as many Seyferts do not show any evidence of starburst activity \citep[e.g.][]{fil93}, and optical
spectroscopic studies of large samples do not indicate that starbursts are more common
in Seyferts than in normal galaxies \citep{pogge89}. In addition, \citet[][hereafter CF04]{cid04} studied 
the optical stellar population (SP) of 65 Sy~2 and 14 other galaxies from the \citet{joguet01} sample. 
They concluded that the star formation history (SFH) of the Sy~2 galaxies is remarkably heterogeneous.
These results are similar to those obtained in the study of the UV and optical SP 
of Seyfert galaxies (mostly Sy~2) available in the literature 
\citep[e.g.][CF04, and references therein]{bica88,schmitt96,cid98,
rosa98,charles2000,raimann03,rosa04,cid05}.

To determine if circumnuclear SPs and nuclear activity are closely 
related phenomena, or if they are only incidental, it is of utmost importance the correct characterisation 
of the former, since a substantial fraction of the energy emitted 
by a galaxy in the optical to NIR domain is starlight. 
Moreover, the analysis of the stellar content provides information on 
critical processes such as the star formation episodes and the 
evolutionary history of the galaxy. 
In this respect, the use of NIR features in the study of SP 
is not recent, dating back to nearly three decades ago. For example,
\citet{rieke80} employed NIR spectroscopy to 
study NGC\,253 and M\,82. They report the detection of a strong 
2.2\mc\ CO band, suggesting the presence of a dominant
population of red giants and supergiants in the nuclear region
of both sources.  Since
then, other authors have used the NIR to study star formation, 
in most cases based on the 2.2\mc\ CO band 
\citep[e.g.][and references therein]{orig93,oliva95,engelbracht98,lancon01}
or photometric methods \citep[e.g.][]{moorwood82,hunt03}.

One reason to use the NIR to study the SP of AGNs is that it is the most convenient
spectral region accessible to ground-based telescopes to probe highly obscured sources. 
However, tracking the star formation in the NIR is complicated \citep{origlia00}. Except for 
a few studies such as those based on the Br$\gamma$ emission or the CO(2-0)
first overtone \citep[e.g.][]{orig93,oliva95}, the SP
of the inner few hundred parsecs of active galaxies in the NIR remains poorly known.
Because stellar absorption features in the NIR are widely believed to
provide a means for recognizing red supergiants \citep{oliva95}, they
arise as prime indicators for tracing starbursts in galaxies.
Besides the  short-lived red supergiants, the NIR also includes 
the contribution of thermally- pulsating asymptotic giant
branch (TP-AGB) stars, enhanced in young to intermediate age stellar
populations \citep[$0.2 \leq t \leq 2$ Gyr,][]{maraston98,maraston05}.
The TP-AGB phase becomes fully developed 
in stars with degenerate carbon oxygen cores \citep[see][for a review]{ir83}.
 Evidence of this population in the optical is usually missed, as the most
prominent spectral features associated with this population fall in the NIR \citep{maraston05}. 

With the new generations of Evolutionary Population Synthesis (EPS) models, which 
include a proper treatment of the TP-AGB phase \citep{maraston05}, it is now possible to study in more 
detail NIR SP of galaxies. According
to these models, the effects of TP-AGB stars in the NIR spectra are unavoidable. 
\citet{maraston05} models, by including empirical 
spectra of oxygen-rich stars \citep{lw00}, are able to foresee the presence of NIR 
absorption features such as the 1.1\mc\ CN band \citep{riffel07}, whose  
detection can be taken as an unambiguous evidence of a young to
intermediate age SP.

Given the above, we feel motivated to carry out the first detailed study of the 
SP in active galaxies in the NIR using the entire 0.8-2.4\mc\ spectral range. 
This paper is structured as follows: The data
are presented in Sect.~\ref{data}. NIR spectral indices are presented in Sect.~\ref{ewobs}.
In Sect.~\ref{sintes} we describe the
fitting method. Results are presented and discussed in Sect.~\ref{results}. The final
remarks are given in Sect.~\ref{finalremarks}.

\section{The Data Set}\label{data}

We chose for this work  a subsample of 24 from the 47 AGNs with NIR spectra
published by \citet{rrp06}. The selected targets display prominent absorption
lines/bands and are listed in Tab~\ref{eqw_galaxias}. 
All spectra were obtained at the NASA 3\,m Infrared Telescope
Facility (IRTF). The SpeX spectrograph \citep{ray03} was used in the
short cross-dispersed mode (SXD, 0.8-2.4\mc). The detector used was a
1024$\times$1024 ALADDIN 3 InSb array with a spatial scale of
0.15''/pixel. A 0.8''$\times$15'' slit was employed giving a spectral
resolution of 360 \kms.  The  radius of the central integrated region is few hundred parsecs for almost 
all sources\footnote{Lower than 300\,pc for 15 objects, between 300\,pc and 500\,pc for 6 and $>$500 for 3. For more details see col. 
10 of Table~1 of \citet{rrp06}}.
For more details on the instrumental
configuration, data reduction, calibration processes and integrated region see \citet{rrp06}. A rapid inspection of
Figs.\,9, 10 and  12 in \citet{rrp06} shows  that 
all the chosen spectra are dominated by strong absorption features, the most prominent 
ones are identified in Fig\,.1 of \citet{riffel08}.

\section{Near infrared spectral indices: direct measurements}\label{ewobs}

For comparison with published works and future NIR stellar population studies, we compute the 
equivalent widths (\w) of the NIR absorption features
as well as selected continuum fluxes (\F) in regions free from emission/absorption, normalized to
unity  at 1.223\mc. \w\ and \F\ are measured according to the 
definitions of \citet{riffel08}. The values of \w\ and \F\ are presented in 
Tabs.~\ref{eqw_galaxias} and \ref{flux_galaxia}, respectively.

\begin{table*}
\renewcommand{\tabcolsep}{.65mm}
\caption[]{Equivalent widths measured in the galaxy sample (in \AA).}
\label{eqw_galaxias}
\begin{scriptsize}
\begin{tabular}{lccccccccccccccc}
\noalign{\smallskip}
\hline
\hline
\noalign{\smallskip}
Object/Ion &CaT$\rm _1$ & CaT$\rm _2$ & CaT$\rm _3$       & CN   & Al\,{\sc i} & Na\,{\sc i} & Si\,{\sc i} & Mg\,{\sc i} &  Si\,{\sc i}   &   CO   &Na\,{\sc i} & Ca\,{\sc i} & CO     & CO     &   CO \\
BP$\rm _b$(\mc)& 0.8476 & 0.8520 & 0.8640 & 1.0780 & 1.1200 & 1.1335 & 1.2025 & 1.5720 & 1.5870 & 1.6110 & 2.1936 & 2.2570 & 2.2860 &  2.3150  & 2.3420 \\
BP$\rm _r$(\mc)& 0.8520 & 0.8564 & 0.8700 & 1.1120 & 1.1300 & 1.1455 & 1.2200 & 1.5830 & 1.5940 & 1.6285 & 2.2150 & 2.2740 & 2.3100 &  2.3360  & 2.3670  \\
\noalign{\smallskip}
center(\mc)& 0.8498 & 0.8542 & 0.8670 & 1.0950 & 1.1250 & 1.1395 & 1.2112 & 1.5771 & 1.5894 & 1.6175 & 2.2063 & 2.2655 & 2.2980 &  2.3255  &  2.3545 \\
 (1)     &  (2)   &  (3)   &   (4)  &  (5)   &  (6)   &  (7)   &   (8)  &  (9)   &  (10)  &  (11)  &  (12)  & (13)   & (14)   &  (15)    & (16)  \\  
\hline
\noalign{\smallskip} 
\multicolumn{16}{c}{Seyfert 2}\\
\noalign{\smallskip} 
\hline
NGC\,262(\dg)   &  3.82\pp0.44 & 5.92\pp0.44 & 4.11\pp0.42 &  - 	  &  -  	&  -		&  -	      &  1.41\pp0.27 &  2.17\pp0.05 &  -	   & 1.04\pp0.04  &  -  	& 2.75\pp0.01 & 0.74\pp0.03 & 1.78\pp0.02	 \\
 Mrk\,993	&  1.41\pp0.15 & 3.46\pp0.13 & 2.73\pp0.12 &  - 	  &  -  	& 2.57 \pp0.01  &  -	      &  3.58\pp0.13 &  2.22\pp0.05 & 2.43\pp0.04  & 3.03\pp0.04  &  -  	& 9.04\pp0.01 & 5.24\pp0.03 & 8.34\pp0.03	 \\
 NGC\,591	&  5.80\pp0.39 & 4.57\pp0.29 & 4.72\pp0.17 & 16.95\pp0.13 &  -  	&  -		&  -	      &  3.83\pp0.07 &  2.47\pp0.02 & 2.81\pp0.03  & 4.51\pp0.21  & 5.39\pp0.13 & 9.83\pp0.11 & 1.39\pp0.03 &10.14\pp0.09	\\
 Mrk\,573	&  3.80\pp0.18 & 4.04\pp0.18 & 3.08\pp0.17 & 14.05\pp0.15 &  -  	&  -		& 2.86\pp0.01 &  3.92\pp0.01 &  2.06\pp0.01 & 2.76\pp0.05  & 2.77\pp0.04  & 1.48\pp0.03 & 7.87\pp0.43 &  -	    & 4.26\pp0.13	  \\
 NGC\,1144	&  4.72\pp0.43 & 5.88\pp0.35 & 4.28\pp0.21 & 18.90\pp0.21 & 2.14\pp0.01 & 3.41 \pp0.01  & 3.30\pp0.01 &  4.27\pp0.04 &  2.31\pp0.13 & 2.81\pp0.01  & 4.01\pp0.02  & 1.70\pp0.09 &10.63\pp0.09 & 5.76\pp0.01 &  -	     \\
 Mrk\,1066	&  4.05\pp0.11 & 4.50\pp0.08 & 3.61\pp0.07 & 13.82\pp0.14 &  -  	& 1.90 \pp0.01  &  -	      &  4.58\pp0.01 &  2.25\pp0.02 & 3.48\pp0.13  & 3.68\pp0.01  & 3.11\pp0.02 & 9.98\pp0.28 & 4.81\pp0.13 & 7.87\pp0.01	  \\
 NGC\,1275	&     -        &   -	     &  -	   &   -	  &    -	& -		&  -	      &       -      &        -     & 2.71\pp0.10  & 3.93\pp0.08  & 2.49\pp0.07 &     -        & -	     &  -	    \\
 NGC\,2110(\dg) &  5.28\pp0.28 & 7.47\pp0.90 & 4.59\pp0.20 & 20.00\pp0.20 &  -  	& 2.28 \pp0.01  &  -	      &  3.78\pp0.05 &  1.81\pp0.01 & 3.20\pp0.28  & 3.30\pp0.04  & 0.75\pp0.01 & 4.78\pp0.05 & 2.32\pp0.10 & 4.20\pp0.04	  \\
 ESO\,428-G014  &  4.31\pp0.06 & 6.30\pp0.04 & 3.85\pp0.01 &  - 	  &  -  	& 2.87 \pp0.07  &  -	      &  4.77\pp0.01 &  2.57\pp0.01 & 3.40\pp0.03  & 4.80\pp0.06  & 2.31\pp0.10 &12.34\pp0.34 & 5.10\pp0.25 &11.80\pp0.11      \\
 Mrk\,1210	&   -	       &  -	     &  -	   &  - 	  &  -  	&  -		&  -	      &  6.54\pp0.19 &  2.82\pp0.16 & 3.35\pp0.13  &  - 	  & 1.07\pp0.16 & 7.73\pp0.15 &  -	    & 3.41\pp0.47	\\
 NGC\,5728	&   -	       &  -	     &  -	   &  - 	  &  -  	&  -		&  -	      &  5.41\pp0.17 &  3.74\pp0.29 & 5.73\pp0.12  & 8.74\pp0.42  & 5.44\pp0.14 & 8.16\pp0.01 & 8.98\pp0.01 &10.27\pp0.01      \\
 NGC\,5929	&  3.56\pp0.19 & 5.58\pp0.17 & 4.06\pp0.15 & 15.42\pp0.25 &  -  	& 1.00 \pp0.91  & 2.17\pp0.91 &  4.27\pp0.49 &  1.70\pp0.19 & 3.52\pp0.21  & 5.43\pp0.17  & 3.67\pp0.36 &13.77\pp0.06 & 6.78\pp0.03 &11.42\pp0.04	   \\
 NGC\,5953(\dg)	&  4.68\pp0.23 & 6.50\pp0.17 & 5.26\pp0.13 & 12.65\pp0.35 & 1.95\pp0.01 & 1.98 \pp0.01  & 1.62\pp0.01 &  3.98\pp0.08 &  1.90\pp0.04 & 3.61\pp0.15  & 3.82\pp0.03  & 2.67\pp0.04 &13.37\pp0.20 & 7.74\pp0.03 &  -	     \\
 NGC\,7674	&  2.24\pp0.51 & 3.42\pp0.43 & 3.57\pp0.31 &  - 	  &  -  	& 3.20 \pp0.36  &  -	      &  2.35\pp0.14 &  1.98\pp0.03 &  -	   &  - 	   &  - 	 &    - 	&  -	     &  -	    \\
 NGC\,7682	&  3.22\pp0.20 & 4.53\pp0.17 & 1.49\pp0.30 & 11.00\pp0.23 & -		&  -		&  -	      &  4.04\pp0.08 &  2.92\pp0.02 & 2.70\pp0.02  &   -	   &  - 	 & 9.42\pp0.26 & 8.33\pp0.08 &  -	     \\
\noalign{\smallskip} 
\hline
\noalign{\smallskip} 
\multicolumn{16}{c}{Seyfert 1}\\
\noalign{\smallskip} 
\hline
Mrk\,334     &   -          &  -          &  -          & 19.11\pp0.10 &  -              &  -            & 1.12\pp0.11 &  3.40\pp0.06   &  1.78\pp0.08  & 3.13\pp0.09  & 1.66\pp0.04  & 1.85\pp0.07  & 6.14\pp0.01  & 2.96\pp0.00  & 3.54\pp0.05   \\
NGC\,1097    &  4.48\pp0.17 & 5.84\pp0.15 & 3.29\pp0.13 & 5.88\pp0.26  & 1.79\pp0.22     & 1.46\pp0.13   & 1.81\pp0.13 &  4.51\pp0.15   &  2.62\pp0.20  & 3.69\pp0.10  & 2.99\pp0.01  & 2.04\pp0.05  & 9.45\pp0.20  & 5.16\pp0.12  &10.32\pp0.02  \\
MCG-5-13-17  &  2.35\pp0.10 & 3.18\pp0.08 & 3.84\pp0.08 &  -	       &  -	         & 1.91\pp0.21   &  -          &  2.87\pp0.04   &  2.25\pp0.05  & 3.07\pp0.06  & 1.43\pp0.05  & 1.97\pp0.02  & 7.95\pp0.11  & 3.87\pp0.03  & 6.55\pp0.05   \\
Mrk\,124     &   -          &  -          &  -          &  -           &  -	         &  -            &  -	       &  2.19\pp0.06   &  1.29\pp0.30  & 2.12\pp0.05  &  -           &  -           &  -           &  -           &  -           \\
NGC\,3227    &   -          & 1.67\pp0.06 & 2.09\pp0.05 &  -           &  -	         & 5.78\pp0.15   &  -	       &  3.43\pp0.03   &  1.58\pp0.03  &  -           & 2.36\pp0.03  & 1.35\pp0.04  & 6.24\pp0.01  & 2.79\pp0.00  & 4.47\pp0.03   \\
NGC\,4051    &   -          &  -          &  -          &  -           &  -	         &  -            &  -	       &  3.22\pp0.01   &  1.93\pp0.03  & 3.19\pp0.08  & 1.13\pp0.01  & 0.73\pp0.01  & 3.69\pp0.11  & 0.66\pp0.08  & 3.50\pp0.02  \\
Mrk\,291     &  -           & 2.40\pp0.07 & 4.45\pp0.09 &  -           &  -	         & 2.56\pp0.01   &  -	       &  3.43\pp0.02   &   -           & 3.80\pp0.11  &  -           &  -           &11.90\pp0.05  &  -           &  -           \\
Arp\,102B    &  -           & -           &  -          &  -           &  -	         &  -            &  -	       &  2.12\pp0.39   &  1.38\pp0.07  & 3.82\pp0.05  & 1.97\pp0.03  & 2.14\pp0.02  & 4.61\pp0.05  & 2.80\pp0.01  &  -           \\
Mrk\,896     &   -          & -           &  -          &  -           &  -	         &  -            &  -	       &  1.69\pp0.14   &  1.24 \pp0.25 & 2.16\pp0.01  & 0.67\pp0.08  & 0.53\pp0.01  & 1.23\pp0.06  & 1.89\pp0.05  &  -           \\
\noalign{\smallskip} 
\hline
\end{tabular}
\begin{list}{Table Notes:}
\item BP$\rm _b$ and BP$\rm _r$ are the blue and red bandpass boundaries. ($\ddagger$)  The continuum of
$W_{\rm CaT }$ is affected by spurious emission. If we use only the points free from emission to
set the continuum we get 3.28\pp0.20, 5.03\pp0.18 and 3.36\pp0.17 for CaT$\rm _1$,  CaT$\rm _2$,  CaT$\rm _3$, 
respectively.
\end{list}
\end{scriptsize}
\end{table*}

The most studied absorption lines  in the literature and present in our spectral range are the Calcium 
triplet lines 
\citep[CaT\footnote{$W_{\rm CaT}$=$W_{CaT1\,0.849\mu m} + W_{CaT2\,0.854\mu m} + W_{CaT3\,0.867\mu m}$},e.g.][and references
therein]{terlevich90,aurea05,vega08}.

 We have four of our objects (Mrk\,573, Mrk\,1066, NGC\,2110 and Mrk\,1210)
in common with \citet{vega08}. For the first 3 objects we measure values larger than those reported by 
them. The difference is due to the different index definitions. While we use those 
of \citet{bica87} and \citet{vega08} use \citet{cenarro01}. The CaT in Mrk\,1210 occurs near the 
detection limit in our spectrum, which thus preclude measurements. \citet{vega08} report a value of 
6.22$\pm$0.48\AA\ for this feature.  In addition, the differences between our measurements and 
those of \citet{vega08} can be related to the fact that they measure $W_{\rm CaT}$ in their 
synthetic spectra, and therefore our measurements represent better the conditions observed in actual galaxy spectra. 
Other explanation for the discrepancies probably lies in the different apertures used in both works. They 
use a slit width of 2.0$''$ \citep{aurea05} while we use a slit 2.5 times narrower \citep{rrp06}.  
\citet{terlevich90} report CaT measurements for NGC\,3227,  NGC\,262 (Mrk\, 348) and NGC\,5953. Our W$_{\rm CaT}$ 
value for the former is consistent with theirs, while for the latter  two  we measure higher values. 
The discrepancy is again due to the different continuum definitions\footnote{ In order to compare our $W_{\rm CaT}$ values with 
those of \citet{terlevich90} we compute the $W_{\rm CaT}$ of our NGC\,5953 spectrum using their continuum definitio. 
The values thus obtained are: $W_{CaT1}$ =1.7\AA\ $ W_{CaT2}$=4.3\AA\ and $W_{CaT3}$=3.8\AA. 
which are very similar to those of \citet{terlevich90}}.  Note that, for NGC\,3227, 
\citet{terlevich90} used an optional blue continuum band at $\lambda$ 8582\AA, wich provides a continuum 
slope very similar in both studies, wich thus explains the similarity in W$_{\rm CaT}$.

\begin{figure}
\includegraphics[scale=0.45]{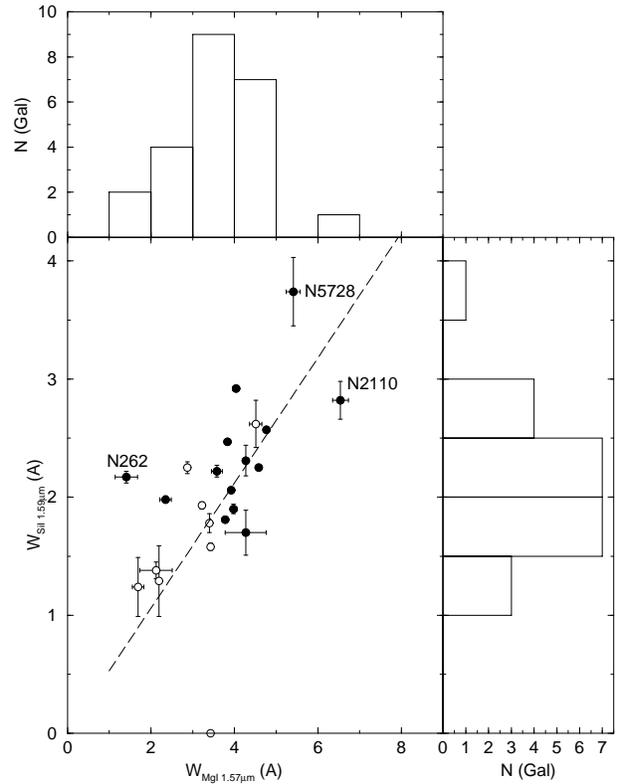}
\caption[]{Diagram of $W_{\rm Mg\,I}$ 1.58\mc\ {\it versus} $W_{\rm Si\,I}$ 1.59\mc. The dashed line is
the a liner correlation. Open symbols are Sy~1 and filled Sy~2.}
\label{figew1}
\end{figure}

In Fig.~\ref{figew1} we compare $W_{\rm Mg\,I}$ 1.58\mc\ with $W_{\rm Si\,I}$ 1.59\mc. These two 
absorption lines are correlated as $W_{\rm Si\,I}$=(0.53$\pm$0.02)$W_{\rm Mg\,I}$, with CC=0.67 (or 
$W_{\rm Si\,I}$=(0.71$\pm$0.21)$W_{\rm Mg\,I}^{(0.8\pm0.2)}$, CC=0.69). The correlation between \w\ of
these two lines suggests that almost all the objects studied here follow the same chemical enrichment. This 
hypothesis can be associated with the fact that Si and Mg are more abundant than the other $\alpha$ 
elements in the Galactic Globular Cluster NGC\,6121, which is located near  
the  Galaxy centre  \citep[$l$=350.97$\rm^o$, $b$= 15.97$\rm^o$,][]{mariano08}.

We have tried diagrams involving other NIR absorption lines (or bands), but only 
weak correlations are found. Since the \F\ values are studied in \citet{rrp06} we only present them 
in Tab.~\ref{flux_galaxia} for comparison purposes. 

\begin{figure}
\centering
\includegraphics[scale=0.45]{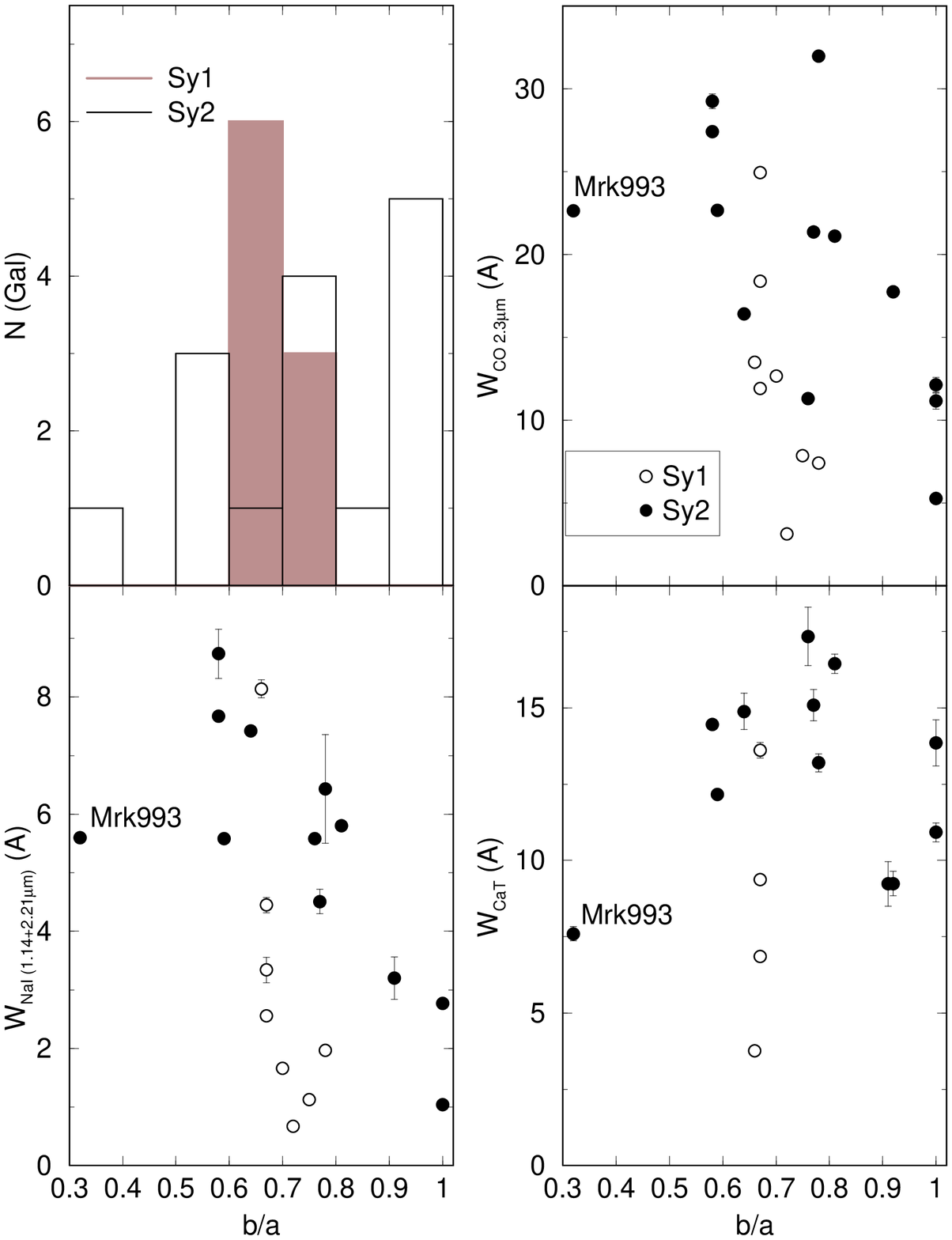}
\caption{Diagrams of \w\ versus b/a. At top left the b/a histogram of our galaxies. The 
other three plots show b/a versus W$\rm _{\rm CO\,2.3\mu m}$, W$\rm _{\rm Na\,I}$ and $W\rm _{CaT}$, respectively. Open
symbols are Sy~1 and filled Sy~2.}
\label{hist_ab} 
\end{figure}

Our sample is composed mostly 
of spiral galaxies (Tab.~\ref{objects}), which tend to increase the
optical $W\rm _{NaI}$ with the inclination b/a \citep[][]{bica91}. This could be 
the case of NGC\,5728, which has a low b/a, high $W\rm _{NaI}$ and displays the highest 
$A\rm _{v}$=3.09\,mag value of our sample. In Fig.\,\ref{hist_ab} we investigate the relation between 
galaxy inclination and \w\ of NIR absorption lines. The correlation 
suggests that the interstellar 
medium plays an important role in the NIR $W\rm _{Na I}$\footnote{
The better NIR correlation can also be associated with the fact that the NIR probes SPs deeper in the 
dust and therefore the contribution of the 
Na\,{\sc i} is enhanced by the intrinsic stellar light.}. On the other hand,  $W\rm _{CO}$ is 
weakly correlated  with  galaxy inclination (Fig.~\ref{hist_ab}, top right), suggesting 
a weaker dependence on the interstellar medium.
No correlation of $W_{\rm CaT}$ with b/a is found. In addition, 
the correlation between Na{\sc i} and $W\rm _{CO}$ 2.3\mc\ with b/a has no evident relation with  
Seyfert type (Fig.~\ref{hist_ab}). 

\begin{table}
\renewcommand{\tabcolsep}{1.60mm}
\centering
\begin{scriptsize}
\caption{Continuum fluxes, normalized to unity at 1.223\mc.}
\label{flux_galaxia}
\begin{tabular}{lccccccccc}
\noalign{\smallskip}
\hline
\hline
\noalign{\smallskip}
Galaxy              & \multicolumn{9}{c}{F$\lambda$/F$_{1.223 \mu m}$}\\
\noalign{\smallskip} 
\cline{2-10}	
\noalign{\smallskip} 	    
                     & 0.81  & 0.88 & 0.99 & 1.06 & 1.22 & 1.52 & 1.70 & 2.09 & 2.19 \\
 (1)                 &  (2)  &  (3) &  (4) &  (5) &  (6) &  (7) &   (8)&  (9) & (10) \\   
\noalign{\smallskip}
\hline
\noalign{\smallskip} 
\multicolumn{10}{c}{Seyfert 2}\\
\noalign{\smallskip} 
\hline
NGC\,262       & 1.28 & 1.12 & 1.11 & 1.01 & 1.00 & 0.94 & 0.95 & 0.95 & 0.92 \\ 
Mrk\,993       & 1.25 & 1.26 & 1.18 & 1.13 & 1.00 & 0.80 & 0.71 & 0.71 & 0.37 \\ 
NGC\,591       & 1.08 & 1.05 & 1.05 & 1.01 & 1.00 & 0.76 & 0.67 & 0.67 & 0.35 \\ 
Mrk\,573       & 1.26 & 1.27 & 1.23 & 1.15 & 1.00 & 0.79 & 0.72 & 0.72 & 0.45 \\ 
NGC\,1144      & 1.25 & 1.25 & 1.21 & 1.16 & 1.00 & 0.76 & 0.67 & 0.67 & 0.34 \\ 
Mrk\,1066      & 1.05 & 1.11 & 1.10 & 1.08 & 1.00 & 0.83 & 0.80 & 0.80 & 0.49 \\ 
NGC\,1275      & 1.47 & 1.29 & 1.20 & 1.16 & 1.00 & 0.87 & 0.81 & 0.73 & 0.73 \\
NGC\,2110      & 0.97 & 0.95 & 0.98 & 1.05 & 1.00 & 0.93 & 0.89 & 0.89 & 0.65 \\ 
ESO\,428-G014  & 1.18 & 1.17 & 1.12 & 1.08 & 1.00 & 0.79 & 0.72 & 0.72 & 0.37 \\ 
Mrk\,1210      &   -  & 1.25 & 1.27 & 1.16 & 1.00 & 0.85 & 0.82 & 0.82 & 0.65 \\ 
NGC\,5728      &   -  &   -  & 1.01 & 1.00 & 1.00 & 0.82 & 0.76 & 0.76 & 0.42 \\ 
NGC\,5929      & 1.25 & 1.24 & 1.15 & 1.10 & 1.00 & 0.80 & 0.73 & 0.73 & 0.35 \\ 
NGC\,5953      & 1.28 & 1.26 & 1.17 & 1.12 & 1.00 & 0.76 & 0.69 & 0.69 & 0.33 \\ 
NGC\,7674      & 1.21 & 1.12 & 1.08 & 1.02 & 1.00 & 0.97 & 1.02 & 1.02 & 1.10 \\ 
NGC\,7682      & 1.04 & 1.06 & 1.07 & 1.05 & 1.00 & 0.83 & 0.74 & 0.74 & 0.35 \\ 
\noalign{\smallskip} 
\hline
\noalign{\smallskip} 
\multicolumn{10}{c}{Seyfert 1}\\
\noalign{\smallskip} 
\hline
Mrk\,334      &  1.07 & 1.07 & 1.08 & 1.06 & 1.00 & 0.93 & 0.91 & 0.91 & 0.69 \\ 
NGC\,1097     &  1.08 & 1.25 & 1.16 & 1.15 & 1.00 & 0.81 & 0.74 & 0.74 & 0.39 \\ 
MCG-5-13-17   &  1.43 & 1.41 & 1.25 & 1.18 & 1.00 & 0.79 & 0.73 & 0.73 & 0.44 \\ 
Mrk\,124      &  -    & -    & 1.00 & 1.00 & 1.00 & 1.07 & 1.06 & 1.06 & 1.00 \\ 
NGC\,3227     &  1.28 & 1.25 & 1.16 & 1.08 & 1.00 & 0.88 & 0.84 & 0.84 & 0.58 \\ 
NGC\,4051     &  -    & 1.27 & 1.15 & 1.07 & 1.00 & 0.89 & 0.86 & 0.86 & 0.73 \\ 
Mrk\,291      &  1.73 & 1.51 & 1.30 & 1.26 & 1.00 & 0.80 & 0.70 & 0.70 & 0.41 \\ 
Arp\,102B     &  1.48 & 1.33 & 1.24 & 1.23 & 1.00 & 0.77 & 0.69 & 0.69 & 0.40 \\ 
Mrk\,896      &  1.31 & 1.20 & 1.12 & 1.07 & 1.00 & 0.95 & 0.93 & 0.93 & 0.73 \\ 
\noalign{\smallskip}
\hline
\noalign{\smallskip}
\end{tabular}
\begin{list}{Table Notes:}
\item The errors on \F\ are $\leq$ 3\% in all cases.
\end{list}
\end{scriptsize}
\end{table}

\section{spectral synthesis}\label{sintes}

In this section we study the NIR SP of our galaxy sample, 
fitting the underlying continuum of the 24 AGNs in the spectral range between 0.8\mc\ and 2.4\mc.

\subsection{The base set\label{base}}

Clearly, the most important ingredient in the SP synthesis is the spectral base 
set, $b_{j,\lambda}$. An ideal base of elements should cover the range of spectral properties observed 
in the galaxy sample, providing enough resolution in age and metallicity to properly address the desired
scientific question \citep[][CF05]{alex91}.

One improvement here over previous approaches that attempted to describe the stellar content
of active galaxies using NIR spectroscopy is the inclusion of EPS models that
take into account the effects of TP-AGB stars. Accordingly, we use as base set the EPS of 
\citet{maraston05}. The SSPs used in this work
cover 12 ages, $t$= 0.01, 0.03, 0.05, 0.1, 0.2, 0.5, 0.7, 1, 2, 5, 9, 13\,Gyr, and 4 metallicities, namely: 
$Z$= 0.02\,$Z_\odot$, 0.5\,$Z_\odot$, 1\,$Z_\odot$ and 2\,$Z_\odot$, summing up 48 SSPs.

When trying to describe the continuum observed in AGNs, the signature
of the central engine cannot be ignored. Usually, this component is represented by a featureless continuum 
\citep[$FC$, e.g.][CF04]{koski78} of power-law form that follows the expression $F_{\nu}\propto \nu^{-1.5}$.
Therefore, this component was also
added to the base of elements. The contribution of this continuum (in percentage)
to the flux at $\lambda _0$ (1.223\mc) is denoted by $FC$ in Tab.~\ref{objects}. According to the unified model 
\citep[e.g.][]{antonucci85,antonucci93}, the $FC$ in Sy~2 galaxies (if present) is due to scattered light
from the hidden Seyfert~1 (Sy~1) nucleus. However, the reader must bear in mind that a common problem
in the study of the SP of Seyfert galaxies (especially in the optical) is that 
a reddened young starburst ($t\leq$5M\,yr) is 
indistinguishable from an AGN-type continuum \citep[see Sec.~\ref{secFC};][CF04]{cid95,thaisa00}. 
To avoid this problem we have not included very young SSPs in our base (this point is also 
discussed in Sec.~\ref{secFC}).

\begin{figure}
\centering
\includegraphics[scale=0.45]{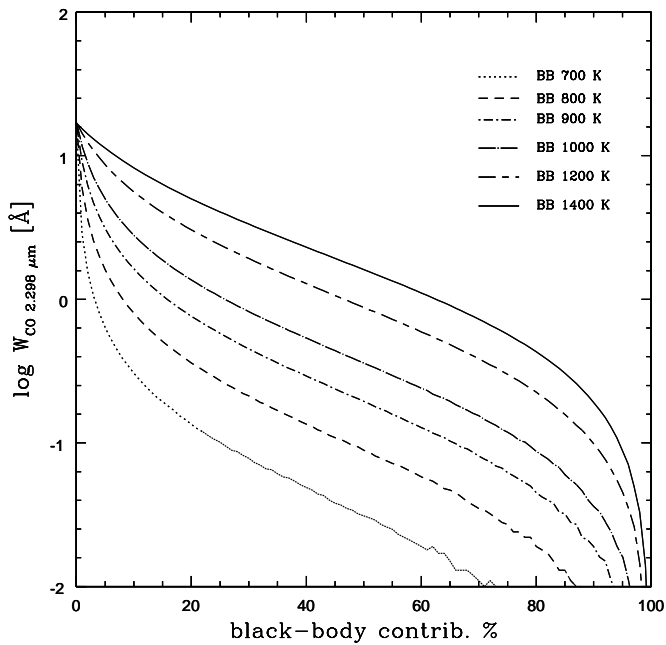}
\caption{Effect on W$_{CO}$ of the combination of the template derived for the starburst galaxy NGC\,7714 with  black-body
distributions at different temperatures. We use increasing $BB$ fractions from 1\% to 100\%. 
$BB$ temperatures are labelled.}
\label{d1} 
\end{figure}

In the spectral region studied here, hot dust plays an important role in the continuum emission of active galaxies.  
Previous studies  \citep[i.e.,][for instance]{rrp06} report a minimum in the continuum emission 
around 1.2\mc, probably 
associated with the red end of the optical continuum related to the central 
engine and the onset of the emission due to reprocessed nuclear radiation by dust 
\citep{bar87,thom95,rudy00,ara03,rom06,rrp06}.  In order to properly account for this component,
we have included in our spectral base 8 Planck distributions (black-body-$BB$), 
with $T$ ranging from 700 to 1400\,K, in steps of 100\,K. The lower limit in $T$ is due to the fact
that lower temperatures are hard to detect. Even a small fractional 
contribution would require a sizeable amount of dust. In order to illustrate this point, we plot 
in Fig.~\ref{d1} combinations of different  $BB$ distributions with 
the synthetic template of the starburst galaxy NGC\,7714, derived by \citet{riffel08}.The
combination was made summing up, in the whole spectral range, increasing fractions of the dust 
component from 0\% to 100\%. Thus, we start with the pure NGC\,7714 synthetic spectrum and 
end with a pure BB distribution according to:
\begin{equation}
F= \left(1-\frac{f}{100}\right)F_{SP} + \left(\frac{f}{100}\right)F_{BB},
\end{equation}
where $f$ is the percentual flux, which we vary in steps of 1\%; $F_{SP}$ is the flux of  the synthetic
spectrum of NGC\,7714 normalized to unity 
at 1.223\mc\ and $F_{BB}$ is the
$BB$ flux also normalized at the same wavelength. As can be 
observed in Fig.~\ref{d1}, small fractional contributions of cool ($<$700\,K) dust can significantly 
alter the strength of the absorption lines of the $K$ band spectrum, and therefore, are very hard to 
be detected in our spectral range. 
Hotter $BB$ distributions are not used because 1400\,K is very close to the sublimation temperature of 
graphite grains \citep[likely the main  constituent of the dust,][]{bar87,rom06}.

\subsection{The method }\label{metod}

The second most important ingredient of a stellar population synthesis
is is the code that will
suitably combine the individual components of the base elements to construct the final model 
that will represent the observed continuum. Here, as a synthesis code, we use for the first time -- in this 
spectral range -- the {\sc starlight} software
\citep{cid04,cid05a,mateus06,asari07,cid08}. This code is well described 
in \citet[CF04,][hereafter CF05]{cid05a}. In summary,
{\sc starlight} mixes computational techniques originally developed for semi empirical 
population synthesis with ingredients from evolutionary synthesis models(CF05). Basically, 
the code fits an observed spectum $O_{\lambda}$ with a combination, in different proportions, of 
$N_{\star}$ single stellar populations (SSPs). Due to the fact that the \citet{maraston05} models
include the effect of the TP-AGB phase, crucial to model NIR SP \citep[see][]{riffel07,riffel08,riffel08c}, 
we used this EPS models as the base set for {\sc starlight}\footnote{As default base set, the code use the 
SSPs of \citet{bc03}.}. 
Extinction is modeled by {\sc starlight} as due to foreground dust, and 
parametrised by the V-band extinction $A_V$. We use the CCM \citep{ccm89} extinction law. 
Essentially, the code solves the following equation for a model spectrum $M_{\lambda}$ (CF05):
\begin{equation}
M_{\lambda}=M_{\lambda 0}\left[\sum_{j=1}^{N_{\star}}x_j\,b_{j,\lambda}\,r_{\lambda} \right] \otimes G(v_{\star},\sigma_{\star})
\end{equation}
were $b_{j,\lambda}\,r_{\lambda}$ is the reddened spectrum of the $j$th SSP normalized at 
$\lambda_0$; $r_{\lambda}=10^{-0.4(A_{\lambda}-A_{\lambda 0})}$ is the reddening term; $M_{\lambda 0}$ is the
synthetic flux at the normalisation wavelength; $\vec{x}$ is the population vector; $\otimes$ denotes the convolution 
operator and $G(v_{\star},\sigma_{\star})$ is the gaussian distribution used to model the line-of-sight stellar
motions, it is centred at velocity $v_{\star}$  with dispersion  $\sigma_{\star}$. 

The fit is carried out with a simulated annealing plus Metropolis scheme, which searches for the 
minimum of the equation (CF05): 

\begin{equation}
\chi^2 = \sum_{\lambda}[(O_{\lambda}-M_{\lambda})w_{\lambda}]^2
\end{equation}
where emission lines and spurious features are masked out by fixing $w_{\lambda}$=0. 
For more details on {\sc starlight} see CF04 and CF05.

\section{Synthesis results and discussion}\label{results}

\begin{figure*}
\centering
\includegraphics[scale=0.9]{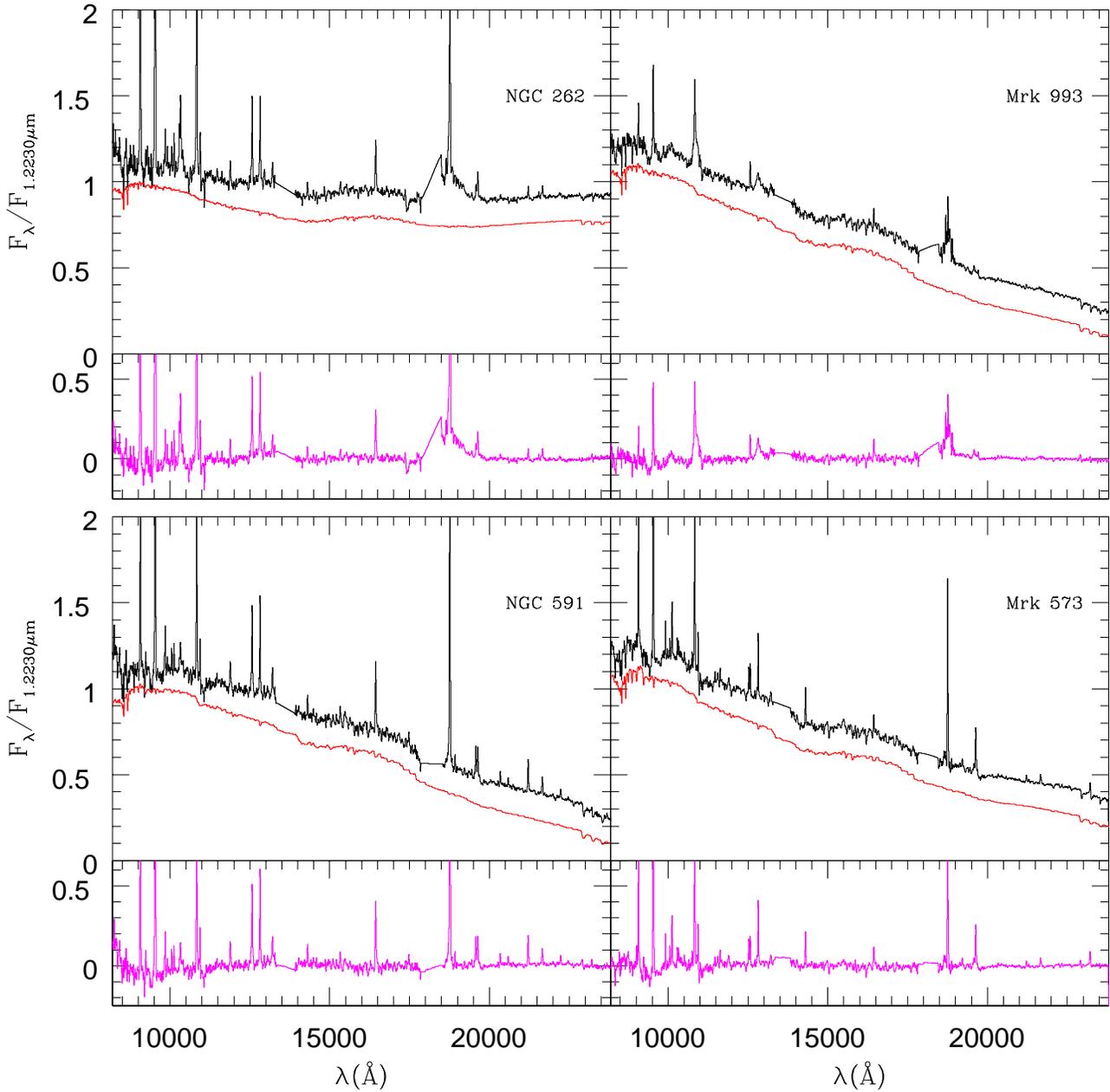}
\caption{Spectral fits. Each panel shows: (i) at top the flux of the observed spectrum,
normalized at unity at 1.223\mc, and the synthetic spectra (shifted down for clarity); (ii) at bottom the 
$O_{\lambda} - M_{\lambda}$ residual spectrum.}
\label{f1}
\end{figure*}

\begin{figure*}
\centering
\includegraphics[scale=0.9]{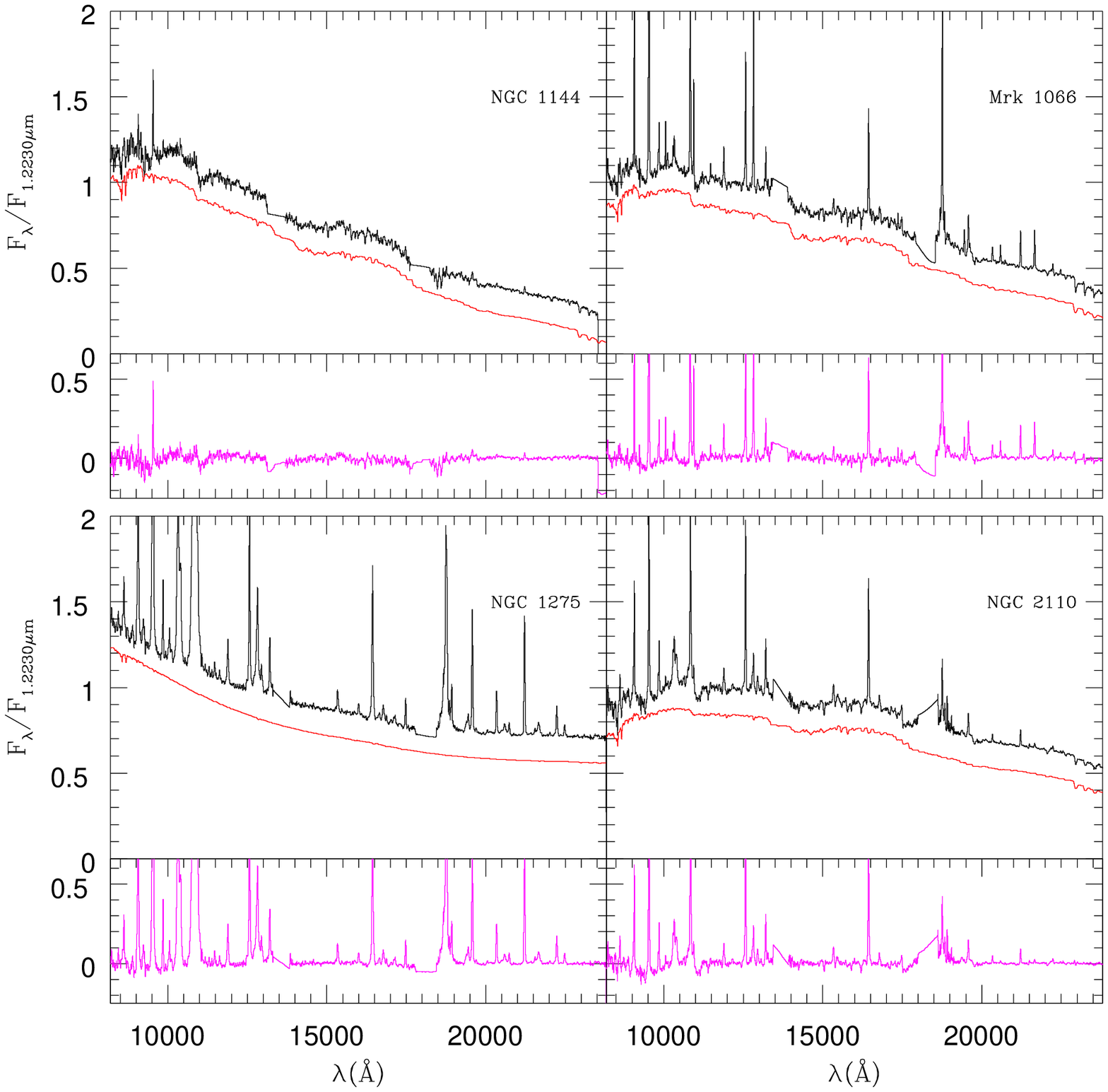}
\caption{Same as Fig.~\ref{f1}}
\label{f2}
\end{figure*}

\begin{figure*}
\centering
\includegraphics[scale=0.9]{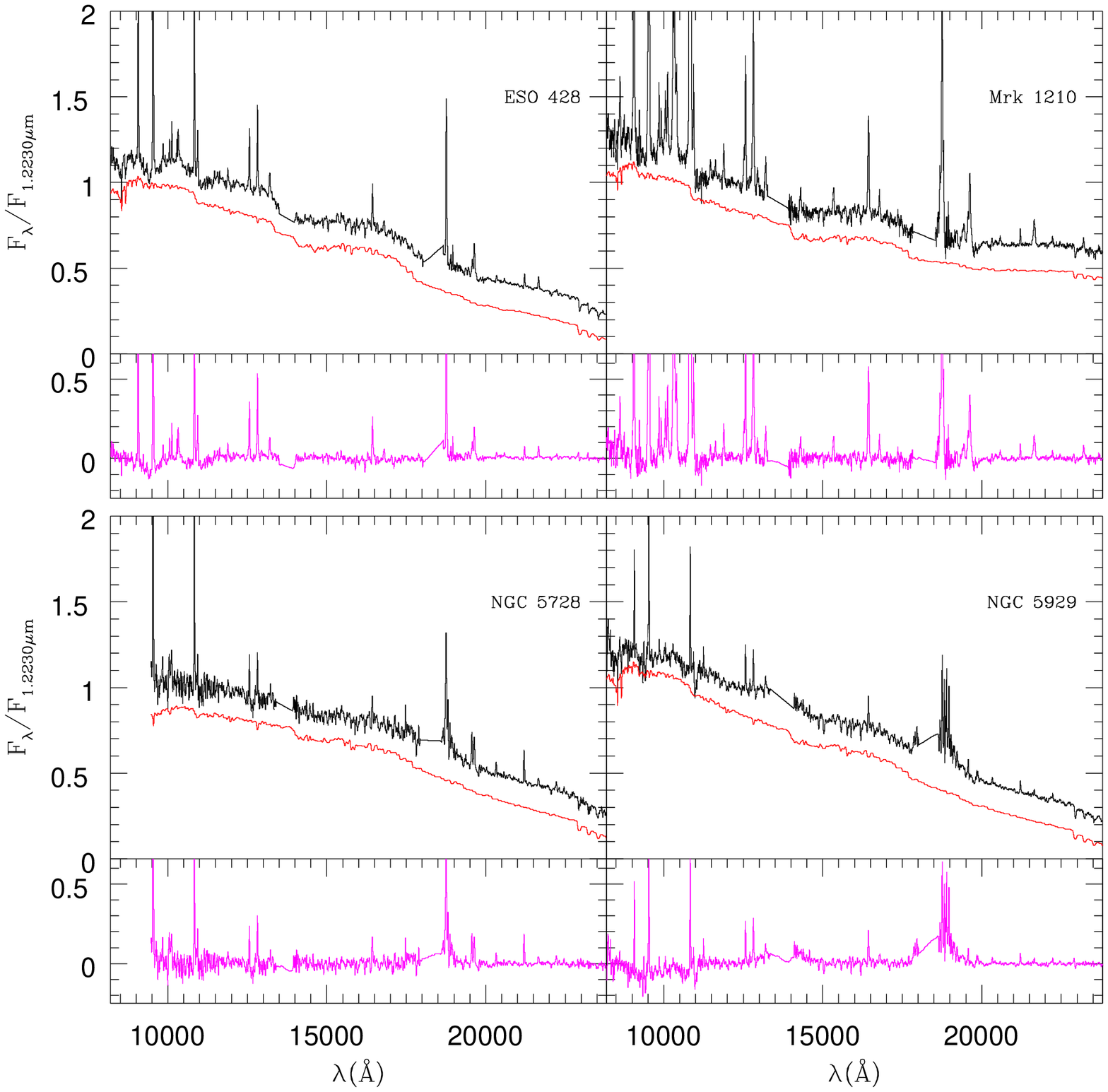}
\caption{Same as Fig.~\ref{f1}}
\label{f3}
\end{figure*}

\begin{figure*}
\centering
\includegraphics[scale=0.9]{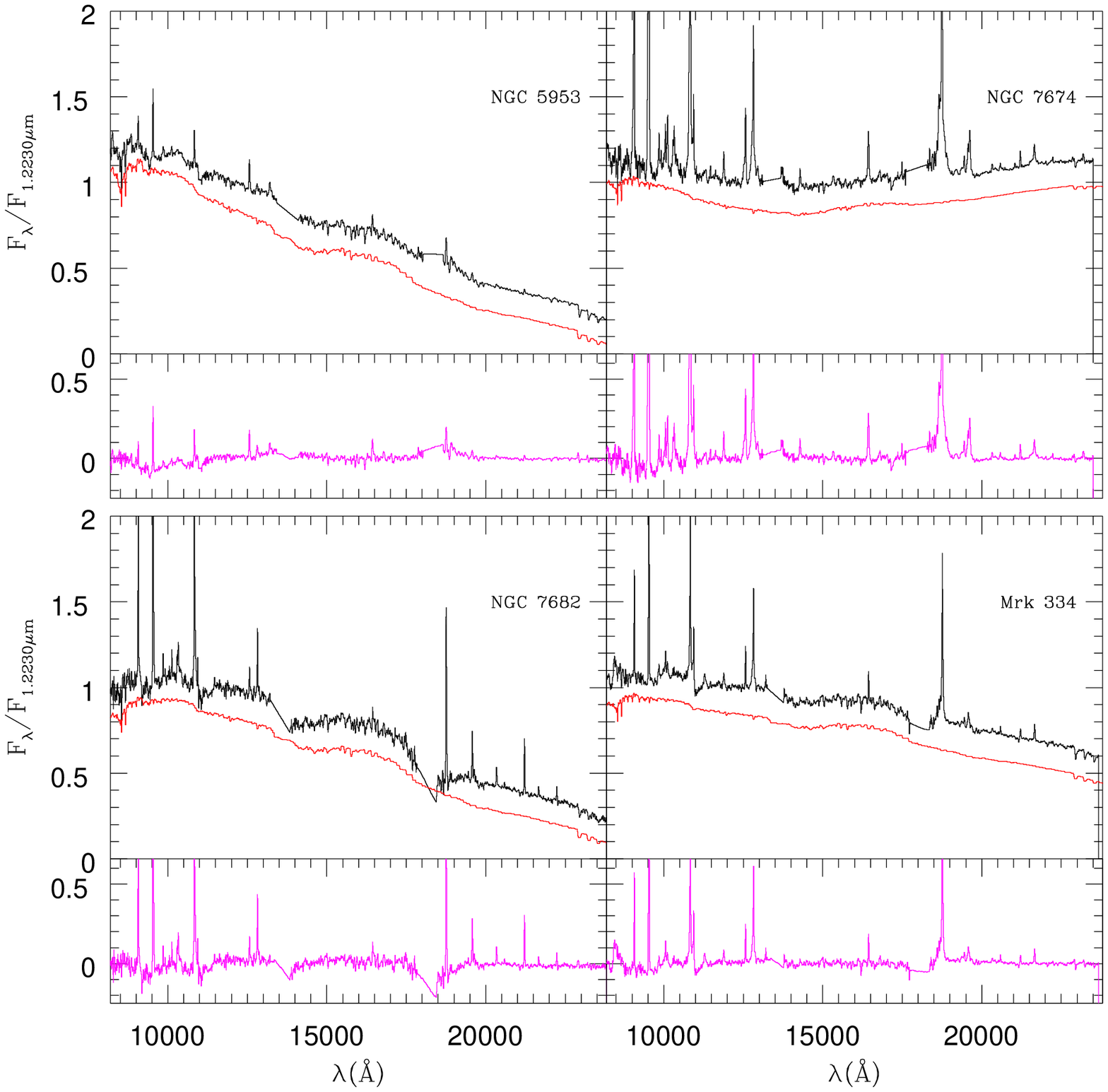}
\caption{Same as Fig.~\ref{f1}}
\label{f4}
\end{figure*}

\begin{figure*}
\centering
\includegraphics[scale=0.9]{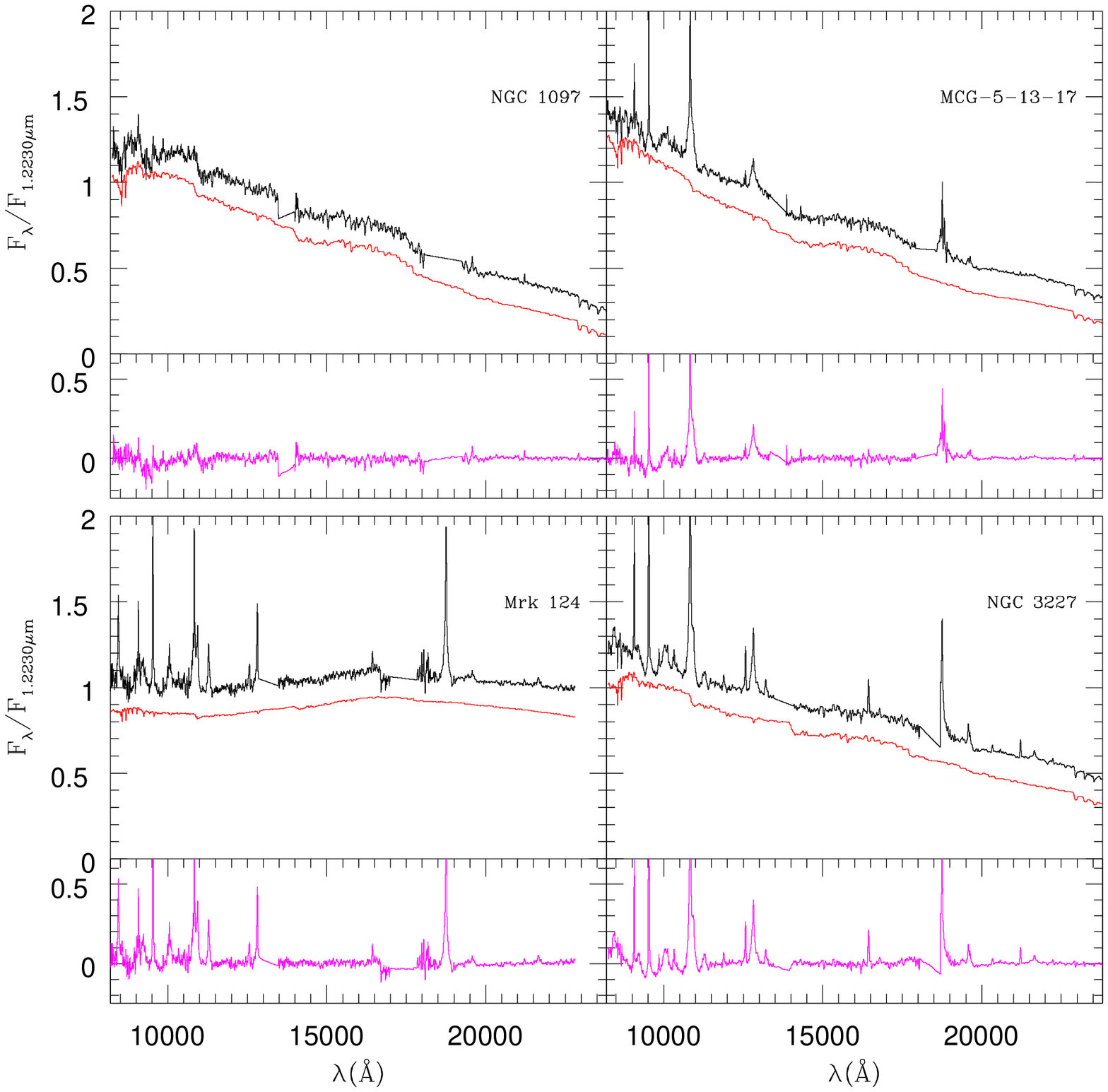}
\caption{Same as Fig.~\ref{f1}}
\label{f5}
\end{figure*}

\begin{figure*}
\centering
\includegraphics[scale=0.9]{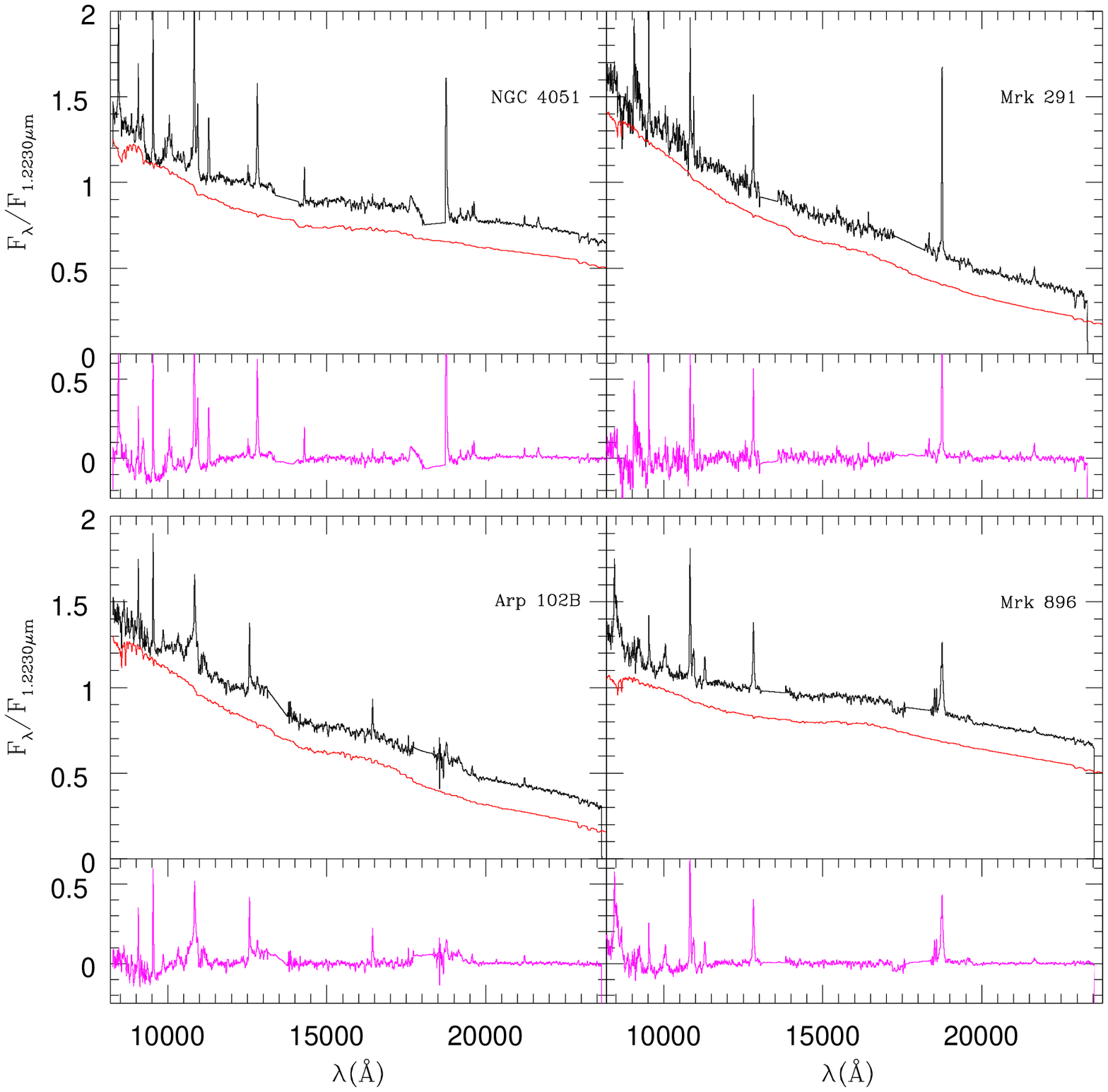}
\caption{Same as Fig.~\ref{f1}}
\label{f6}
\end{figure*}

We present in Figs.~\ref{f1} to \ref{f6} the results of the spectral synthesis fitting procedure. 
For each galaxy the top panel shows the observed and synthetic spectra normalized to unity at 1.223\mc.
Note that in all cases the 
synthetic spectrum was shifted by a constant for visualisation purposes. The bottom panel 
shows the residual spectrum $O_{\lambda} - M_{\lambda}$. As expected, the residual
is dominated by the nebular emission. The analysis of the emission lines free from the SP 
contamination is beyond the scope of this paper and is left for a forthcoming work
(Riffel et al., 2009 {\it in preparation}).  The results of the synthesis are 
summarised in Tab.~\ref{objects}. The quality of the fits are measured by the $\chi^2$ (column 13)\footnote{ 
Which, in fact, is the $\chi^2$ divided by the number of $\lambda$'s used in the fit. 
Reliable fits are obtained when $\chi^2\sim$1 (CF04). for more details
see {\sc starlight} manual available at http://www.starlight.ufsc.br.} and the {\it adev} 
(column 14) parameters. The latter gives 
the percentage mean deviation $|O_{\lambda}- M_{\lambda}|/O_{\lambda}$ over all fitted pixels.

\begin{table*}
\begin{small}
\renewcommand{\tabcolsep}{0.70mm}
\caption{Synthesis Results. For more details see text.}  
\label{objects}  
\begin{tabular}{lccccccccccccccccccc}
\hline\hline
\noalign{\smallskip}
Galaxy		&b/a$^\ddag$   & Morphology$^\dag$& $FC$  &  $BB_c$   &  $BB_h$ & $x_y$ &  $x_I$  &  $x_O$ & $m_Y$ & $m_i$  &  $m_O$   & $\chi^2$  & adev	  &  Av   & $\langle {\rm log}\,\,t_\star \rangle _L$ & $\langle {\rm log}\,\,t_\star \rangle _M$ & $\langle Z\star \rangle _L$ & $\langle Z_\star \rangle _M$  \\
                &              &             &  (\%) &   (\%)    &  (\%)   &  (\%) &  (\%)   &  (\%)  &  (\%) &  (\%)  &  (\%)    &	       &	  & (mag) &  (yr)				      & (yr)					 		& ($\ast$)		     &	($\ast$)  \\
(1)             &      (2)     & (3)   &   (4)     &   (5)   &  (6)  &   (7)   &  (8)   &  (9)  &  (10)  &  (11)    &  (12)     &  (13) & (14) &   (15)                                    &            (16)                           &   (17)     &(18) &(19)            \\
\noalign{\smallskip}
\hline 
\hline
\noalign{\smallskip} 
\multicolumn{19}{c}{Seyfert 2}\\
\noalign{\smallskip} 
\hline
NGC\,262      & 1.0  & S0-a	&   20.0   & 1.8   & 3.6   &  0.0   &   13.0  &   61.2  &   0.0   &   2.9   &   97.1  &   1.52  &  1.06  &  1.37 &  9.68    &  9.89     &  0.011 & 0.005 \\
Mrk\,993      & 0.32 & Sa	&   4.7    & 0.0   & 0.0   &  0.0   &	27.8  &   66.8  &   0.0   &   6.4   &	93.6  &   0.01  &  1.90  &  1.32 &  9.54    &	9.83	&  0.013 & 0.005\\
NGC\,591      & 0.77 & S0-a	&   0.0    & 0.0   & 0.0   &  14.9  &	27.4  &   57.5  &   1.4   &   7.5   &	91.1  &   0.01  &  2.87  &  1.97 &  9.28    &	9.84	&  0.014 & 0.003\\
Mrk\,573      & 1.0  & S0-a	&   22.4   & 0.0   & 0.0   &  11.9  &	52.6  &   11.8  &   3.4   &   45.1  &	51.5  &   2.98  &  1.02  &  0.99 &  8.93    &	9.56	&  0.028 & 0.024\\
NGC\,1144     & 0.64 & E	&   0.0    & 0.0   & 0.0   &  0.0   &	71.3  &   26.1  &   0.0   &   44.9  &	55.1  &   0.01  &  2.22  &  1.44 &  8.97    &	9.32	&  0.021 & 0.015\\
Mrk\,1066     & 0.59 & S0-a	&   17.9   & 0.0   & 0.0   &  4.6   &   50.7  &   26.8  &   0.8   &   29.8  &   69.5  &   2.36  &  1.11  &  1.54 &  9.19    &   9.67    &  0.025 & 0.015 \\
NGC\,1275     & 0.77 & S0	&   65.8   & 0.3   & 0.0   &  2.8   &   0.0   &   32.1  &   0.1   &   0.0   &   99.9  &   1.15  &  0.80  &  0.22 &  9.68    &   9.94    &  0.003 & 0.001 \\
NGC\,2110     & 0.76 & E-SO	&   33.6   & 0.0   & 0.9   &  9.5   &   36.3  &   18.9  &   1.7   &   26.0  &   72.3  &   1.42  &  0.75  &  1.98 &  8.98    &   9.70    &  0.028 & 0.010 \\
ESO\,428-G014 & 0.58 & S0	&   0.0    & 0.0   & 0.0   &  9.2   &	49.1  &   39.6  &   1.1   &   15.9  &	83.0  &   0.03  &  2.05  &  1.52 &  9.13    &	9.76	&  0.021 & 0.005\\
Mrk\,1210     & 1.0  & S?	&   12.0   & 0.5   & 0.0   &  9.2   &   48.8  &   29.1  &   0.3   &   30.7  &   69.0  &   2.29  &  1.05  &  1.14 &  9.17    &   9.77    &  0.013 & 0.007\\
NGC\,5728     & 0.58 & Sa	&   0.0    & 0.0   & 0.0   &  0.0   &	69.5  &   27.8  &   0.0   &   24.4  &	75.6  &   0.01  &  2.89  &  3.09 &  8.64    &	9.51	&  0.037 & 0.037\\
NGC\,5929     & 0.78 & Sa	&   0.0    & 0.0   & 0.0   &  7.3   &	24.5  &   69.7  &   1.4   &   11.0  &	87.7  &   0.01  &  2.38  &  1.48 &  9.37    &	9.61	&  0.020 & 0.022\\
NGC\,5953     & 0.81 & S0-a	&   0.0    & 0.0   & 0.0   &  0.0   &	33.5  &   65.2  &   0.0   &   7.0   &	93.0  &   0.01  &  2.00  &  0.81 &  9.46    &	9.80	&  0.031 & 0.023\\
NGC\,7674     & 0.91 & SBbc	&   29.2   & 2.2   & 5.7   &  3.4   &	24.6  &   35.6  &   0.7   &   15.8  &	83.4  &   0.03  &  1.37  &  1.03 &  9.28    &	9.69	&  0.018 & 0.010\\
NGC\,7682     & 0.92 & Sab	&   0.0    & 0.0   & 0.0   &  0.0   &	34.7  &   62.2  &   0.0   &   6.4   &	93.6  &   0.02  &  3.76  &  1.89 &  9.53    &	9.91	&  0.020 & 0.011\\
\noalign{\smallskip} 
\hline
\noalign{\smallskip} 
\multicolumn{19}{c}{Seyfert 1}\\
\noalign{\smallskip} 
\hline
Mrk\,334      &  0.70 & Sbc	 &   23.7   & 0.0   & 8.2   &  12.2  &   24.1  &   31.7  &   2.6   &   9.9   &   87.5  &   0.01  &  1.45  &  1.36 &  9.03    &	 9.71	 &  0.018 & 0.007 \\
NGC\,1097     &  0.67 & SBb	 &   4.3    & 0.0   & 0.0   &  19.8  &   46.4  &   29.8  &   2.9   &   28.6  &   68.5  &   1.70  &  1.33  &  1.29 &  8.88    &   9.49    &  0.024 & 0.017\\
MCG-5-13-17   &  0.67 & E-SO     &   6.6    & 0.0   & 3.9   &  22.9  &   45.7  &   21.6  &   5.2   &   25.1  &   69.7  &   1.68  &  0.81  &  0.88 &  8.55    &	 9.47	 &  0.023 & 0.012 \\
Mrk\,124      &  0.67 & S?	 &   36.5   & 0.0   & 22.6  &  3.6   &   18.3  &   19.0  &   1.4   &   14.8  &   83.8  &   0.10  &  1.29  &  0.81 &  9.10    &	 9.73	 &  0.008 & 0.002 \\
NGC\,3227     &  0.66 & SABa     &   31.0   & 0.0   & 2.1   &  38.4  &   28.1  &   0.0   &   10.8  &   89.2  &   0.0   &   0.05  &  1.24  &  0.99 &  8.00    &	 9.07	 &  0.011 & 0.011 \\
NGC\,4051     &  0.75 & SABb     &   41.0   & 0.0   & 6.8   &  37.6  &   14.4  &   0.0   &   46.7  &   53.1  &   0.2   &   0.03  &  1.28  &  0.57 &  7.53    &	 8.02	 &  0.020 & 0.010 \\
Mrk\,291      &  0.67 & SBa	 &   6.5    & 0.0   & 4.5   &  0.0   &   6.1   &   85.3  &   0.0   &   1.1   &   98.9  &   1.01  &  2.06  &  0.45 &  9.85    &	 9.94	 &  0.006 & 0.004 \\
Arp\,102\,B   &  0.78 & ?	 &   7.9    & 0.0   & 3.0   &  11.7  &   35.2  &   41.5  &   1.0   &   12.7  &   86.3  &   1.61  &  1.14  &  0.68 &  9.08    &	 9.80	 &  0.015 & 0.005 \\
Mrk\,896      &  0.72 &  Sab     &   27.0   & 0.0   & 13.7  &  4.5   &   2.3   &   53.0  &   0.5   &   0.6   &   98.9  &   0.01  &  0.90  &  0.55 &  9.72    &	 9.96	 &  0.009 & 0.012 \\
\hline
\end{tabular}
\begin{list}{Table notes:}
\item  ($^\ddag$) From NED. ($^\dag$) From HyperLeda - Database for physics of galaxies \citep[http://leda.univ-lyon1.fr, ][]{paturel03}. 
($\ast$) Abundance by mass with Z$\odot$ = 0.02.
\end{list}

\end{small}
\end{table*}

\begin{figure*}
\begin{minipage}[b]{0.5\linewidth}
\includegraphics[width=\textwidth]{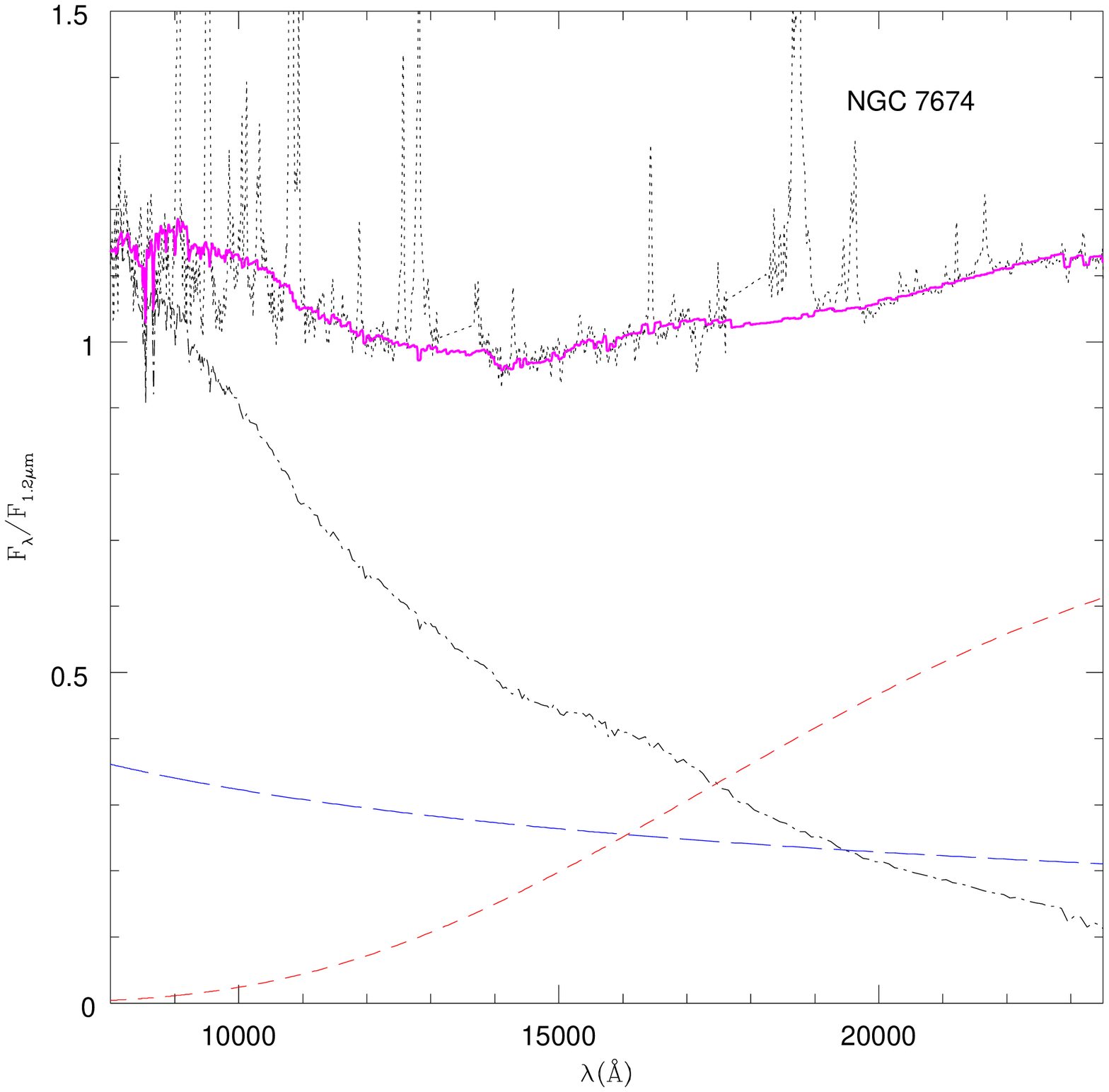}
\end{minipage}\hfill
\begin{minipage}[b]{0.5\linewidth}
\includegraphics[width=\textwidth]{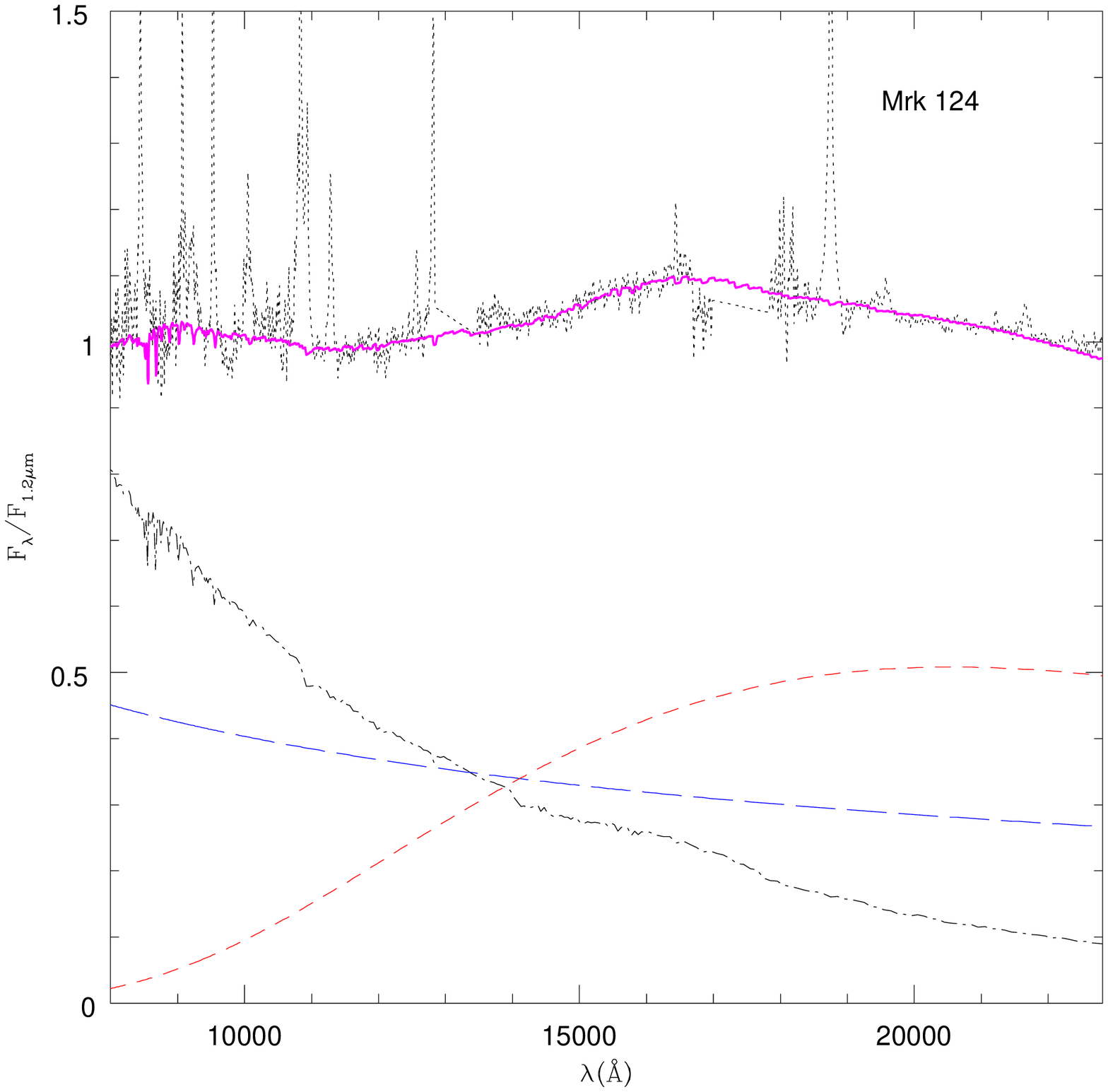}
\end{minipage}\hfill
\caption[]{Three continuum components of NGC\,7674 and Mrk\,124. Dot-short dashed line represents the SP 
($x_{\rm y} + x_{\rm I} + x_{\rm o})$. The FC and hot dust component are represented 
by the long and short dashed lines, respectively. The solid line is the sum of the three components and the 
dotted line represents the observed spectrum.}
\label{3comp}
\end{figure*}

To take into account noise effects that dump small differences between 
similar spectral components, we followed CF05 and present our 
results using a condensed population vector, which is obtained by binning
the $\vec{x}$ into {\bf young}, $x_Y$ ($t_j\leq \rm 5\times 10^7$yr); {\bf intermediate-age}, 
$x_I$ ($1\times 10^8 \leq t_j\leq \rm 2\times 10^9$yr) and {\bf old}, $x_O$ ($t_j > \rm 2\times 10^9$yr) components, 
using the flux contributions. The same bins were used to represent the mass components of the population vector; 
$m_Y$, $m_I$ and $m_O$, respectively. The condensed population vectors are presented in columns 7 to 12 of
Tab.~\ref{objects}.
For more details on vector definition see CF05.  We have also binned the black-body 
contributions into two components. The cool ($BB_c$) is obtained by summing 
up the $BB$ contributions with T $\leq$ 1000\,K, and  the hot one ($BB_h$) with T$\geq$ 1100\,K. These components
were defined based on the sublimation temperatures of silicate ($\sim$1000\,K) and graphite ($\sim$ 1200\,K) 
grains \citep{bar87,granto94}. The condensed black-body vectors for each source are listed in columns 5 and 6 of
Tab.~\ref{objects}.

The spectral synthesis shows that the NIR continuum of active galaxies 
can be explained in terms of at least three components: 
a non-thermal continuum, the dust emission and the SP of the circumnuclear region. As can be seen
in Table~\ref{objects} and in Fig.~\ref{3comp},  the contribution of the latter to the nuclear continuum
is higher than 50\% in most objects.  Therefore, its study is a critical step in the analysis of the continuum 
emission of Seyfert galaxies.   Moreover, our results are consistent with the predictions of the unified model for AGNs, 
as the non-thermal continuum and the hot dust emission are present in all Sy\,1 sources and only in a 
small fraction of the Sy~2s (see also Sec.~\ref{secdust}).
In the following sections we provide a detailed description of each of these three components 
in the light of the results obtained here and compare them with those obtained in 
other wavelength regions, mostly in the optical.

\subsection{The stellar population}

\begin{figure*}
\begin{minipage}[b]{0.45\linewidth}
\includegraphics[scale=0.45]{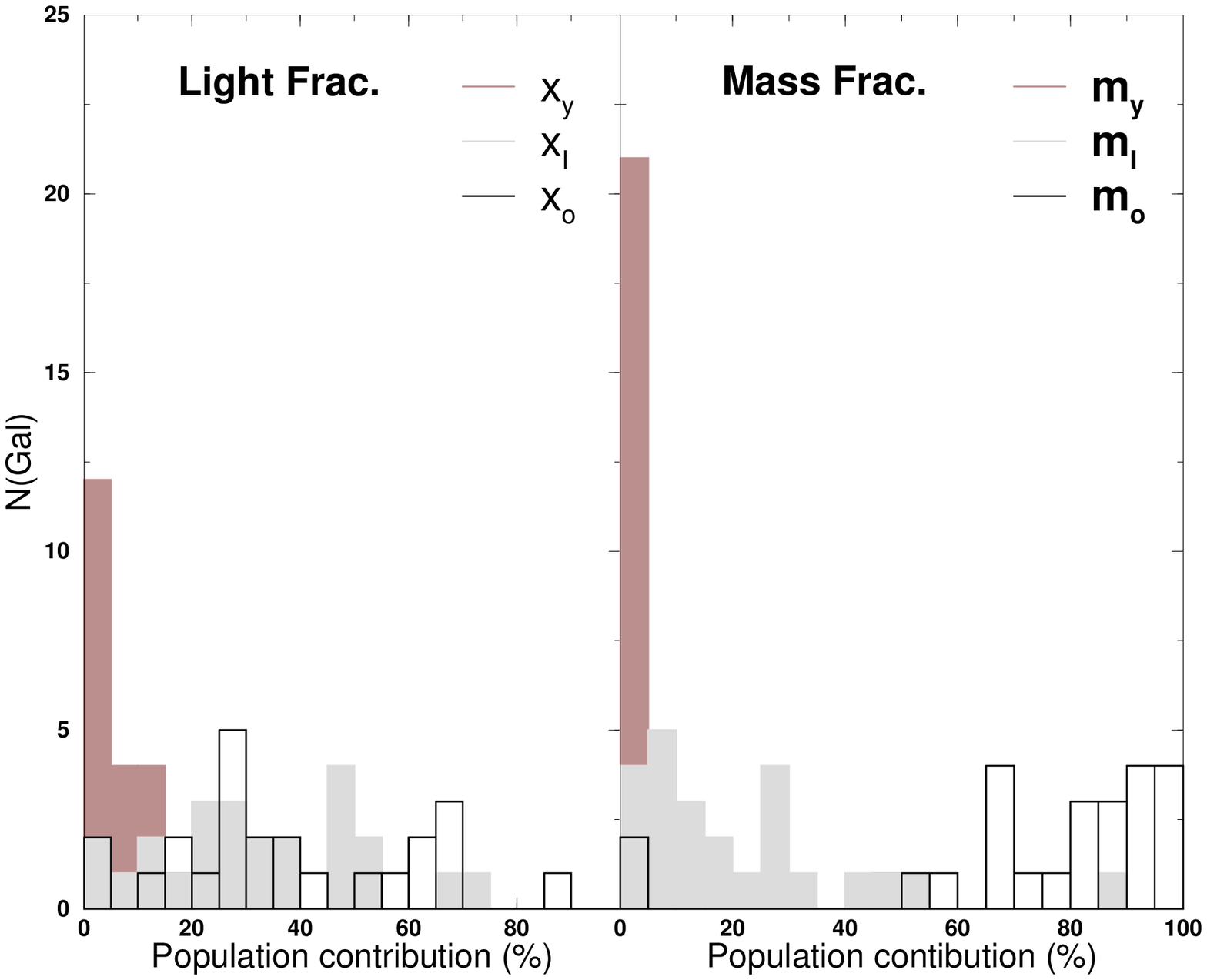}
\caption{Histograms comparing  the population vector components. Light-weighted 
at the left panel and mass-weighted to the right one. The lines (colours) indicating each component of $\rm \vec{v}$ are on the labels.}
\label{hist_vectors}
\end{minipage}
\hfill
\begin{minipage}[b]{0.45\linewidth}
\includegraphics[scale=0.35, angle=-90]{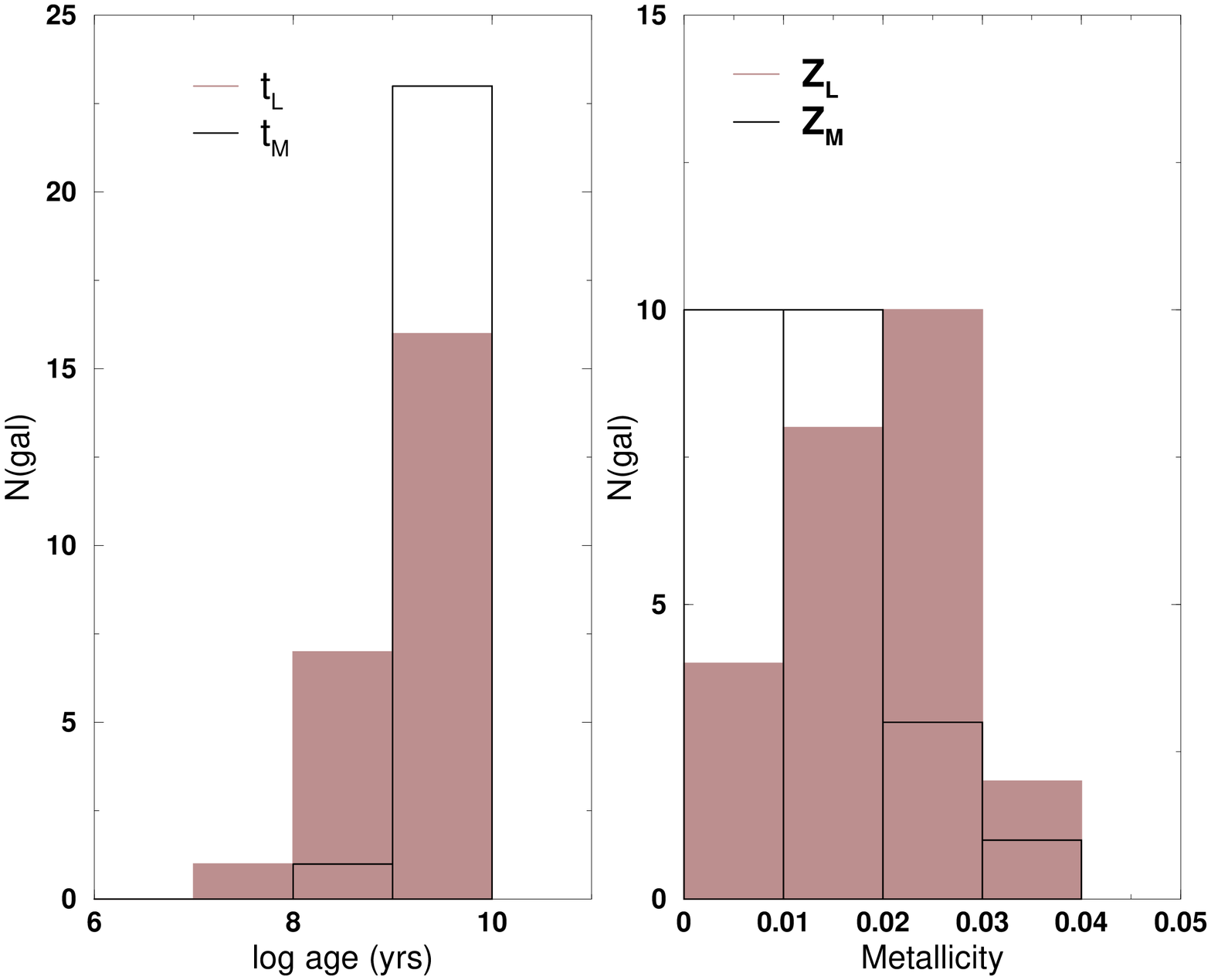}
\caption{Histograms comparing our results in terms of light 
and mass fractions. At left  we show a 
histogram of light- (filled) and mass-weighted (solid line) mean ages. 
At the right we present a histogram comparing the light (filled) 
and  mass-weighted mean metallicities.}
\label{histam}
\end{minipage}
\hfill
\end{figure*}

Binning the population vectors into six components\footnote{With 3 components 
representing the star formation episodes.} left us with a 
coarser but more powerful description of the SFH of our galaxy sample. To better quantify the NIR SFH 
we plot a histogram with the flux-weighted and mass-weighted 
condensed population vectors in Fig.~\ref{hist_vectors}. Overall, 
the NIR SPs are heterogeneous, as  in most sources the three
components contribute significantly to the integrated flux. 
However, the contribution of $x_Y$ is very small ($\lesssim$10\%) for
most of our sample. The intermediate-age component is well distributed, with a maximum 
centred at $\sim$40\%. The $x_O$ contribution is very similar to that of the intermediate age.

Regarding the mass-weighted components,  as expected from the results seen above, 
the contribution of $m_y$ is very 
small (near zero) as can be observed 
in the right side of  Fig.~\ref{hist_vectors}. The intermediate-age mass contributions are distributed over
all fractions (from 0 to 100\%) but tend to be biased to values 
lower than 20\%. In contrast, the old component of the mass-weighted vector is biased to values 
higher than 70\%.

According to CF05, if one would characterise the SP mixture of a galaxy by a single parameter,
it is the mean stellar age. They defined it in two ways: the first is weighted by light fraction,
\begin{equation}\langle {\rm log} t_{\star} \rangle_{L} = \displaystyle \sum^{N_{\star}}_{j=1} x_j {\rm log}t_j, \end{equation}
and the second, weighted by the stellar mass, 
\begin{equation}\langle {\rm log} t_{\star} \rangle_{M} = \displaystyle \sum^{N_{\star}}_{j=1} m_j {\rm log}t_j. \end{equation}
Note that both definitions are limited by the age range used in our elements base (Sect.~\ref{base}) 
and obviously, the $FC$ and $BB$ components were excluded from the sum. The mean 
stellar ages derived with both definitions are presented in columns 16 and 17 of Tab.~\ref{objects}, respectively.

To better quantify the NIR mean ages we show at the left side of Fig.~\ref{histam} 
histograms for $\langle {\rm log} t_{\star} \rangle_{L}$ and $\langle {\rm log} t_{\star} \rangle_{M}$. 
 The light-weighted mean age of our galaxy sample is biased to 
an intermediate/old age SP, while for the mass-weighted mean age
we clearly observe that the old population dominates.
As stated by CF05 the mass-weighted mean age is a more physical 
parameter, but it has a much less direct relation with the observables. They associated 
this discrepancy with the non-constant stellar $M/L$ ratio.

A secondary parameter to describe the mixed SP is the metallicity. CF05 also
defined the light-weighted mean metallicity by 
\begin{equation}\langle Z_{\star} \rangle_{L} = \displaystyle \sum^{N_{\star}}_{j=1} x_j Z_j,\end{equation}
as well as the mass-weighted mean metallicity, which is represented by:
\begin{equation}\langle Z_{\star} \rangle_{M} = \displaystyle \sum^{N_{\star}}_{j=1} m_j Z_j.\end{equation} 
Both definitions are bounded by the $\frac{1}{50}Z_{\odot}$-2$Z_{\odot}$ range. The light- and mass-weighted 
mean metallicities, estimated for our galaxies, are presented in columns 18 and 19 of 
Tab.~\ref{objects}, respectively. 

We present in the right side of Fig.~\ref{histam} a histogram for the 
light- and mass-weighted metallicities of our sample. Our results point 
to a mean metallicity solar to above solar, if we consider the  
light-weighted values,  while for the mass-weighted mean metallicity 
our results indicate a sub-solar value. We associate this discrepancy with the well 
known age-metallicity degeneracy, i.e. for a fixed mass, a high-metallicity SP looks cooler - and older - 
than a low-metallicity SP, thus resulting in a higher $M/L$ ratio.  Moreover, this is consistent 
with a galaxy chemical enrichment scenario in which the young population is enriched by the evolution of the 
early massive stars. In this context, the light-weighted metallicity is more sensitive to the 
young component, while the mass-weighted metallicity to the old stellar population.

\subsubsection{Previous studies\label{previous}}

A significant fraction (12 out of 24) of our objects have been subject of previous
SP studies in the optical and UV regions. In this section we will compare our NIR results 
with those available in the literature.

\begin{itemize}

\item {{\bf NGC\,262:}}  Accordingly to \citet{garcia-vargas89}
most of the optical and NIR flux comes from the galaxy bulge SP. The emission at 
shorter wavelengths can be fitted by a power law. 
 \citet{rosa01} studied the optical SP of this galaxy and 
found a dominant old population. By fitting the nuclear continuum with an off-nuclear 
spectrum of the galaxy bulge as template,  they conclude that the 
contribution of a power law is unnecessary. In the case of an elliptical as a template spectrum, they needed to
include a contribution $\leq$25\% of a power law ($F_{\lambda} \propto \lambda^{-1}$). They argue that 
such small dilution can be related to a change in the SP of NGC\,262 with respect to the 
elliptical galaxy template. Similar results are obtained by \citet{raimann03}. They 
found 60\%, 20\%, 12\% and 8\% for the old, intermediate, young and $FC$ components, respectively\footnote{ Note 
that we have used only the nuclear region. We use the 10\,Gyr, 1\,Gyr, 100+10\,Myr and 3\,Myr+$FC$ 
as old, intermediate, young and $FC$ components, respectively.}. As can be observed in Tab.~\ref{objects} our results 
agree with the optical studies.

\item{{\bf ESO\,428-G014:}} CF04 carried out a study of the SP, in the spectral 
region between 3500 and 5200\,\AA\ of the inner 200 pc of this source. Their study also used
{\sc starlight}. They found $x_y$=20\%, $x_I$=47\% and $x_O$=33\%. As can be observed 
in Tab~\ref{objects} our NIR results agree with those obtained by CF04 in the optical.

\item{{\bf Mrk\,1066:}} The light in the optical region of this galaxy is dominated 
by young to intermediate age SPs, and the nuclear spectrum is strongly diluted by a non-thermal component
\citep{rosa01}.  These results agree with those reported by \citet{raimann03}.
Our results show that light at 1.22\mc\ is dominated by a intermediate age SP diluted 
by a $FC$ component, in agreement with the optical results (see Tab.~\ref{objects}). 
With respect to the stellar component, our results agree with \citet{ramos09}, wich studied the same spectral region
by modeling the continuum with combinations of stars and black-body (with T=1000K) dilution, without the $FC$ component.

\item{{\bf Mrk\,1210:}} The optical SP of this source has been studied by many
authors \citep[][CF04]{schmitt99,thaisa00} however, the results are controversial. While 
\citet{schmitt99} argue that the optical light is dominated by a 10 Gyr population (54\% 
at $\lambda$ 5870\AA), \citet{thaisa00} suggest that a young SP 
(or a power law) contributes with 50\% of the flux observed in $\lambda$4020\AA. When modelling the
spectral interval between $\lambda$3500-5200~\AA, CF04 inferred that 53\% of the
flux observed in the spectrum  Mrk\,1210 was due to a young stellar population. They also found a contribution of
39\% for the old component and 5\% for the intermediate one.  Our results disagree with those reported 
in the optical, as we found a lower fraction for the old component ($\sim$29\%) and 
$\sim$49\% for the intermediate one. Moreover, $\sim$20\% of the flux belongs 
to young+$FC$ components. 

\item{{\bf NGC\,3227:}} We found a dominant young component ($\sim$ 40\%) for this galaxy. Our results 
agrees with those of \citet{davies07,davies06} which analyze the star formation in the inner 10 pc of this galaxy  using 
the near infrared adaptive optics integral field spectrograph SINFONI. They found that the 
light at this region is dominated by a 40\,Myr SP. 

\item{{\bf Mrk\,573:}} The optical SP of this source was studied by \citet{schmitt99}. They fit the 
\w\ and continuum ratios, finding that 82\% of the flux observed in $\lambda$ 5870\AA\ is 
due to a 10 Gyr stellar population. These results are in good agreement with 
those obtained by  \citet{raimann03}, \citet{thaisa00} and \citet{rosa01} in the optical 
region\footnote{These two latter groups have analysed the absorption lines species present in the observed spectra.}. 
Our NIR spectral synthesis points to a dominant intermediate age population ($\sim$53\%) diluted by a $FC$ component,
which contributes with $\sim$22\%  of the observed continuum flux. As for Mrk\,1066, we agree with \citet{ramos09} with respect to the dominant presence of an 
intermediate age population, but we only found dilution by the $FC$ component.

\item{{\bf NGC\,1097:}} The SP of this galaxy was studied in the UV by \citet{charles98}. 
They infer that the light at $\lambda$2646 \AA\ is due to a fraction of $\sim$40\% of a young SP\footnote{We call 
young SP the sum of the contributions with age $\leq$0.2Gyr. For more details see Tab~11 of \citet{charles98}.}, 
16\% to intermediate age and $\sim$44\% to old SP. These results agree with the detection 
of a young starburst in the inner 9 pc of this galaxy \citep[][see also \citet{davies07,davies09}]{thaisa05} and with the  optical SP studied by
CF04, who found a contribution of 31\% for the young component. However, there are some differences 
between the UV and optical SP for the intermediate age (CF found 37\%) and  
for the old (CF found 12\%, they also
determine 19\% for the $FC$) components. The results for the young SP of both groups of authors are  
consistent with our NIR synthesis. Regarding the intermediate and old age components, our results 
are more consistent with those of the optical region. The discrepancies between our results and 
those obtained in the UV are probably related to the fact that in the NIR we are integrating light
through a deeper line of sight and thus, we detect old stars located 
more internally in the bulge of the galaxy.

\item{{\bf NGC\,2110:}} \citet{rosa01} argue that the stellar optical absorption  lines are 
similar to those of an old population. By fitting the nuclear continuum with an off-nuclear 
spectrum of the galaxy bulge as template,  they conclude that the 
contribution of a power law is unnecessary. In the case of an elliptical as a template spectrum, they needed to
include a contribution $\leq$25\% of a power law ($F_{\lambda} \propto \lambda^{-1}$).  
These results are further confirmed by CF04, who found a contribution of 67\% for the old component, 
7\% for the intermediate age and 26\% for the young SP, as well as by \citet{raimann03} who 
determine contributions of, 53\%, 32\% and 10\% for the old, intermediate and young age components, 
respectively. Our results indicate a dominant intermediate age stellar population 
and a strong $FC$ component. The detection
of the latter component may be related to the fact that the NIR is less affected by  
dust extinction and is consistent with the detection of broad components 
in the emission lines. (see Sec. \ref{secFC}).

\item{{\bf NGC\,5728:}} CF04 fitted the SP of this object and concluded that its 
light between 3500 and 5200\,\AA\ is dominated by intermediate age stars (51\%), with  
contributions of 20\% and 28\% of young and old SPs, respectively. These fractions agree with 
our NIR analysis, where we found a dominant contribution of the intermediate age SP ($\sim$70\%) and $\sim$30\%
for the old component. However, we did not detect the young component in our fitting process. 
This can be associated to the fact that NIR light is dominated by intermediate 
age stars \citep{maraston05} or to our spectral coverage that misses the calcium 
triplet absorption, lines which are more sensitive to young SPs \citep[see Fig.~4 of][]{riffel08}.

\item{{\bf NGC\,5929:}} According to \citet{rosa01} the nucleus of this source is dominated by old SP,
a conclusion  obtained by  studying the \w\  of optical absorption lines. They also conclude that
if the off-nuclear spectrum is used as template to fit the nuclear spectrum of NGC\,5929, no 
signs of dilution are observed.  Simmilar results are obtained by \citet{raimann03} who found 
contributions of, 61\%, 27\%, 8\% and 4\% for the old, intermediate and young and $FC$ components, 
respectively. 
These results fully match our NIR synthesis ($FC$=0\%, $x_y$=7.7\%, $x_I$=24.5\% and $x_O$=69.7\%) 

\item{{\bf NGC\,5953:}} One of the galaxies studied by CF04. They derived 
a dominant intermediate age optical SP ($x_I$=74\%) with small fractions of young and old stars ($x_y$=1\% and
$x_O$=7\%) and a contribution of 18\% of the $FC$. The NIR SP of this galaxy is divided into two components, 
the old, dominant, ($\sim$65\%) and the intermediate age ($\sim$34\%). The FC component does not
contribute to the NIR flux.

\item{{\bf NGC\,7682:}}  The optical light of this source is dominated by an old SP (86\%) with no 
contribution of intermediate  age stars (CF04). A small contribution of young populations (8\%)
and 6\% of a $FC$ component is reported by CF04. Our NIR results disagree with those of the optical: 
the SP is  shared between old ($\sim$63\%) and intermediate 
($\sim$35\%) age populations.

\end{itemize}

\begin{figure}
\centering
\includegraphics[scale=0.45]{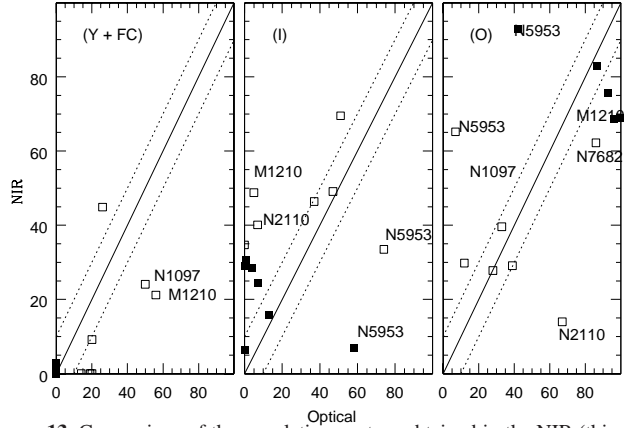}
\vspace{-4cm}
\caption{Comparison of the population vectors obtained in the NIR (this work) and in the optical (CF04), for 7 objects in common.
The symbols indicate the population vectors. The full line is the identity line, the dotted lines represent $\pm$10\% 
deviation from the identity. Open and filled symbols are the flux and mass fraction, respectively.  }
\label{compara}
\end{figure}

Our NIR synthesis and that in the optical (CF04) have been analysed by the same method (the
{\sc starlight} code), allowing for a proper comparison of 
 the objects in common to both studies. Fig.~\ref{compara} 
summarises this comparison. In general, our results do not agree with those in the optical.
Part of the differences can be accounted for by the fact that the NIR is  more suitable for 
the identification of old SPs and the detection of the unique absorptions 
related to 1\,Gyr-old SP featured by TP-AGB
stars \citep{riffel08}. This hypothesis is appropriate for the case of NGC\,5953, where
the differences between our results and those of CF04 occur in the $x_I$ and $x_O$ components. 
In addition, the differences between optical and NIR SPs can be associated to the fact 
that the NIR probes SPs buried deep in the dust. An example of 
this situation is NGC\,2110, where we have detected the presence of hot dust in its integrated spectrum
(Tab.~\ref{dust}). For the case of NGC\,1097, \citet{thaisa05} report an obscuration of a 
central starburst, which they associate with a dusty absorbing medium. 

Our finding that the central regions of the galaxies
contain a substantial fraction of intermediate-age SPs (see Figs.~\ref{hist_vectors} and \ref{cn}), together 
with the prolonged SFH, is very similar to the picture drawn in the case of central 
star-forming rings, based on optical data \citep{allard06,sarzi07}. This might support the scenario
where central star formation often occurs in circumnuclear rings \citep{sarzi05,shields07}.

\subsubsection{The CN {\it versus} intermediate age stellar population}

\begin{figure}
\centering
\includegraphics[scale=0.5]{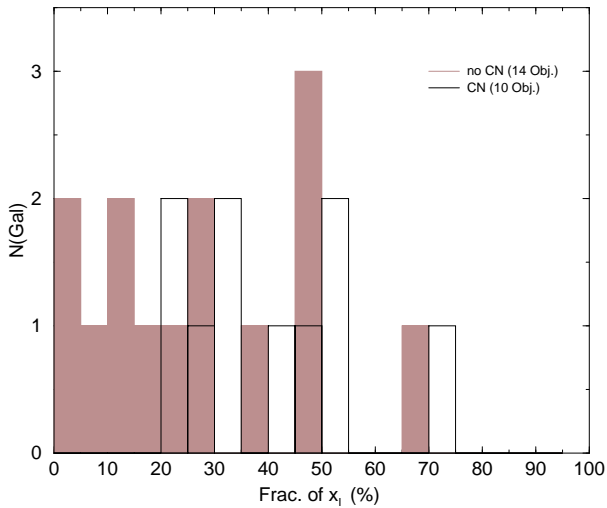}
\caption{Histograms comparing the intermediate age component of the galaxies with CN detection (empty histogram) 
and un-detection (shaded).}
\label{cn}
\end{figure}

Our NIR approach offers a unique opportunity to investigate in a more consistent way 
the relation between the CN molecular band and the unambiguous 
evidence of an intermediate age SP \citep{maraston05,riffel07,riffel08c}. Fig.~\ref{cn} 
presents a histogram comparing the 
intermediate age SP of the galaxies where CN was clearly detected \citep{riffel07} and those
with no detections in a visual inspection. The objects with a
clear CN detection have contributions of the intermediate age component higher than $\sim$20\%, with 
a mean value of  40$\pm$15\%. Regarding the galaxies where CN was not detected, the values are 
biased to  contributions lower than $\sim$30\%. The mean value derived for the intermediate age for 
the sources without CN detection is 27$\pm$20\%.

In four galaxies where CN was not detected we obtained fractions of the $x_I$ component 
higher than $\sim$35\%.  This result may appear contradictory, as the CN band is the
stellar feature that most suitably traces the intermediate age component. For NGC\,5728, for instance, we 
associate the 69\% of the intermediate age component with our shorter spectral coverage 
(see Sec.~\ref{previous}) and  the low S/N ratio.  For ESO\,428-G014, Mrk\,1210 and MCG\,5-13-17, 
the CN band is totally filled by the Pa$\gamma$ emission line \citep[Fig\,2 of][]{rrp06}.

We conclude this section arguing that our NIR spectral fitting confirms that the detection 
of the CN band in the spectrum of a galaxy can be taken 
as an unambiguous evidence of the presence of an intermediate age SP.

\subsection{The featureless component \label{secFC}}

As discussed in Sec~\ref{base}, it is very difficult to distinguish a reddened young strarburst from
a $F_{\nu}\propto \nu^{-1.5}$ power law. However, this effect is even harder 
in the optical, where the main difference between a 5 Myr SSP and a $FC$ seen through an absorption
$A\rm _v \sim 2-3$ mag by dust is the presence of the Balmer absorption lines 
and Balmer jump in the blue side in the former (CF04).

\citet{cid95} predicted that a broad component in H$\beta$ 	
becomes distinguishable whenever the scattered $FC$ contributes with $\geq$20\% to the 
optical continuum light. With our synthesis we can investigate this issue in the NIR. 
Six out of 15 Sy~2 galaxies of our sample display 
a $FC$ contribution higher than $\gtrsim$20\% (Tab.~\ref{objects}). Interestingly, 
a broad component in the Hydrogen lines is detected in the spectra of the 
six sources: Mrk\,1066 \citep[e.g.][]{veilleux97}, Mrk\,573 \citep{nag04}, 
NGC\,1275 \citep[e.g.][]{ho97}, 
NGC\,2110 \citep[e.g.][]{reun03}, NGC\,262 \citep[e.g.][]{miller90} and NGC\,7674 
\citep[e.g.][]{miller90}. 

For two of our galaxies a broad component is reported in the literature,
and we do not detect strong contribution of the $FC$ component. The first one is Mrk\,1210, where 
a broad component of H$\alpha$ and H$\beta$ is detected in polarised 
light \citep{tran92,tran95} and in the NIR H{\sc i} lines \citep{ximena07} for which we find a $\sim$12\% 
contribution of the $FC$ component and $x_Y\sim$10\%. Our NIR synthesis 
for this galaxy is consistent with that obtained by CF04
in the optical, but we tend to find a lower contribution of the 
young component. The second object  is NGC\,5953, where \citet{goncalves99} report the possible detection of 
a very weak broad H$\alpha$ line, but we did not find any contributions 
of $FC$ and $x_Y$. CF04 report a contribution of 18\% for the non-thermal 
component. No broad components are detected in our NIR spectrum \citep[see][]{rrp06}.

The minimum $FC$ contribution predicted by \citet{cid95} seems to be reflected in the NIR, as 
we clearly detect $FC$ fractions higher than 20\%  in the Sy~2 objects with a broad component 
in the H\,{\sc i} lines.
Therefore,  our results reinforce their predictions. In addition, we detect $FC>$20\% in almost all Sy~1 
sources\footnote{With a mean value of $\sim$20\% if all Sy~1 sources are considered and $\sim$ 
32\% if we exclude Mrk\,291, Arp102B, MCG-5-13-17 and NGC\,1097.}, which is consistent with the above arguments. 
However in four of our Sy\,1 galaxies (Mrk\,291, Arp 102\,B, MCG-5-13-17 and NGC\,1097)
we detect fractions of $FC<$20\%.  For the first three objects we associate this 
ambiguity to the absence of features suitable to properly fit the absorption spectra and to a poor S/N ratio in the blue side of the
spectrum (see below). 
As discussed by CF04 and CF05, a high S/N ratio is required for an adequate detection of the different 
contributions to the integrated spectra. Interestingly, the three objects display the presence of hot dust,
wich is in full agreement with the nature of the Sy~1 objects predicted by the unified model for AGN 
(see Sec~\ref{secdust}). 
NGC\,1097,  originally classified as LINER on the basis 
of its optical spectrum \citep{kee83} was reclassified by \citet{thaisa93,thaisa97,thaisa03} 
as a Sy~1 after observing broad Balmer emission lines, a featureless blue continuum 
and double peak profiles. Our spectral synthesis for NGC\,1097 agrees with 
that obtained by CF04. Moreover, our value of $x_Y\sim$20\% is consistent with the starburst 
nature of this galaxy \citep{thaisa05}.

It is worth to mention at this point that the W$_{\rm CaT}$ values, measured for almost 
all object of our sample (see Sec.~\ref{ewobs}), even if the dilution is considered, 
are consitent with values measured in normal spiral galaxies \citep[$\lesssim$15\AA][]{bica87}. 
However, in two cases, NGC\,262 and NGC\,2110, if we account for the dilution we reach very high 
values for W$_{\rm CaT}$ ($\sim$ 17 \AA\ and $\sim$25\AA, respectively). One explanation 
is that the W$_{\rm CaT}$ of these sources is overestimated due to the effects on 
the continuum and absorption lines by telluric features (see Tab.~\ref{eqw_galaxias}).
Other posibility  lies in the fact that the $FC$ component is more sensitive to shorter wavelengths 
($\lambda \lesssim$ 10\,000\AA\ see Fig.~\ref{3comp}), which is our region with 
the lower number of constraints\footnote{Due to the large number of emission lines which 
are masked out.} and the lower S/N (the border of the spectrum; $\lambda \leq$ 9\,000 \AA). 
Therefore, the $FC$ component can be overestimated in these sources.

\subsection{Dust emission}\label{secdust}

A rapid look at column 5 of Table~\ref{objects} shows an excess of emission over the stellar 
population and the featureless continuum for half of the AGN
spectra. This excess (see Fig.~\ref{3comp}), characterised by  
a Planck distribution, suggests emission from hot dust grains. 
Since \citet{bar87}, evidence of the presence of dust near the sublimation 
temperature has been observed 
in the central region of AGNs \citep[e.g.][]{marcoall98,marcoall00,ara05,rom06,rogemar08b}. 
In order to quantify the dust contribution, we show in Tab.~\ref{dust} the individual 
contribution of each blackbody over the temperature interval 800 $-$ 1400~K. Note that
as the T=700\,K and T=1000\,K components were not detected, we dropped them from Tab.~\ref{dust}. 

Only in five Sy\,2 galaxies we have detected hot dust, while a positive detection is reported for all Sy\,1s. 
The obvious reason for this is that we are dealing with hot dust in the $K$-band ($T \sim$1\,000\,K), 
and in the case of Sy\,2s the dust is cooler ($T\sim$600\,K), thus more visible in the $L$ or $M$-bands. 
This hypothesis is supported by the fact that in two Sy\,1 galaxies, Mrk\,1239 and Mrk\,766, we can see directly 
the presence of hot dust in the $K$-band \citep{ara05,rom06}. 
There are two remarkable cases, Mrk\,573 and Mrk\,1066, where we have detected a significant 
fraction of the $FC$ component and no hot dust. There are some 
possible explanations for this, the first two are that discussed above. Another 
possibility is that at least a significant fraction of the detected $FC$ 
component is due to a very young starbust ($t\lesssim$5\,Myr).
The latter, as discussed above (see Sec.~\ref{secFC}) lies in the fact that the 
$FC$ can be overestimated due the small number of constraints in the shorter wavelengths.

The two main constituents of interstellar dust are graphite (carbon) and silicate 
grains \citep[e.g.][]{krugel03}. As discussed in Sect.~\ref{results}, the evaporation temperatures 
of graphite and silicate grains are 1500\,K and $\sim$ 1000\,K, respectively \citep{bar87,granto94}.
The temperature derived for the $BB$ component of almost all objects is $\geq$900\,K 
(see $BB_c$ and $BB_h$ in Tab.~\ref{objects}), suggesting that hot
dust close to the central source is probably composed by graphite grains instead of silicates. The 
only discrepant objects are Mrk\,1210 and NGC\,1275 with T=800\,K. Considering, however, that 
the spatial resolution in those objects is $\sim$ 400 pc \citep{rrp06}, and that the 
dust temperature is a function of the distance to the central source \citep{marcoall98}, it is 
very likely that dust at higher temperatures exists closer to the central source. This would 
rule out the possibility of silicates as the main component of the nuclear dust grains. 

Assuming that the temperature of the different $BB$, found for each galaxy 
represents the distribution of temperatures for graphite grains in the nuclear
region, we can estimate
the lower limit of the hot dust mass responsible for the observed $K$-band excess (see Fig.~\ref{3comp})
following the  approach developed by \citet{bar87}.

 The infrared spectral
luminosity of each dust grain can be obtained from \citep{bar87}:
\begin{equation}
L_{\nu,ir}^{gr}=4\pi^2 a^2Q_{\nu}B_{\nu}(T_{gr}) \rm \ [erg\,s^{-1}\,Hz^{-1}],
\label{lumgr}
\end{equation}
where $a$ is the grain radius; $Q_{\nu}=q_{ir}\nu^{\gamma}$ is its absorption efficiency 
and $B_{\nu}(T_{gr})$ is the Planck function for a grain at temperature $T_{gr}$.
$L_{\nu,ir}^{gr}$ was calculated assuming a typical grain radius $a\rm =0.05\mu m$ \citep{bar87,kishimoto07} 
$q_{ir}$=1.4$\times 10^{-24}$ and $\gamma$=1.6 \citep{bar87}. The values of $L_{\nu,ir}^{gr}$ 
are shown in Tab.~\ref{dust}.

The hot dust mass, M$\rm_{HD}$, can be obtained by the equation \citep{ara05}:
\begin{equation}
M_{HD}\approxeq \frac{4\pi}{3}a^3N_{HD}\rm\rho_{gr}, 
\label{hd}
\end{equation}
where $N_{HD}\approxeq\frac{L_{ir}^{HD}}{L_{\nu,ir}^{gr}}$ is the number of hot dust grains and $\rm \rho_{gr}$
is the density of the grain. $L_{ir}^{HD}$ is the total NIR luminosity due to hot dust. 
It can be derived from the integrated the flux of each $BB$ contribution over the 
spectral range between  0.01 and 160\mc\ found on each galaxy\footnote{
Note that we have weighted the distribution according to its contribution to the total SED.} 
Then, we multiplied the integrated normalized flux by the actual flux at 1.2\mc\ (our normalisation point)
and convert it to the adequate units (from \erg$\AA^{-1}$ to \ergh). The final result of this process 
is the flux of each $BB$ contribution ($F_{BB}$). The values derived for our sample are presented in 
Tab.~\ref{dust}. The $L_{ir}^{HD}$ was estimated using $F_{BB}$ and c$z$ listed in Tab.~\ref{dust} (we have 
adopted $\rm H_0=75\ km\,s^{-1} Mpc^{-1}$).

Finally, we have estimated the lower limit for hot dust mass, for graphite 
grains with $\rm \rho_{gr}=2.26\,g\,cm^{-3}$ \citep{granto94} and using Eq.~\ref{hd}. The hot dust mass of each $BB$ 
distribution, as well as the total hot dust mass ($\sum$M$_{HD}$), are presented in Tab.~\ref{dust}.

\begin{table*}
\begin{small}
\renewcommand{\tabcolsep}{0.60mm}
\caption{Dust properties for our galaxy sample.}  
\label{dust}  
\begin{tabular}{lcrcccccccccccccc}
\hline\hline
\noalign{\smallskip}
       &         &           & \multicolumn{5}{c}{Seyfert 2} &\multicolumn{8}{c}{Seyfert 1} \\
\cline{4-8}
\cline{10-17}       
$T$(K) &$L_{\nu,ir}^{gr}$(a) &    & NGC\,262   &  NGC\,2110&   Mrk\,1210  &   Mrk\,1275      &NGC\,7674 &&   Mrk\,334 &  MCG\,5-13-17  &   Mrk\,124 & NGC\,3227  & NGC\,4051&	 Mrk\,291  &	 Arp\,102\,B &   Mrk\,896  \\ 
       &         &  Cont.(\%) 	  &   0        &  0	   &    0.5	  &	 0.3	     &  0	&&   0	      &  0	       &  0	    &  0	 &  0	   &   0 	  & 0		    &	0	  \\ 
       &         &  F$_{BB}$ (b)  &  	       &	   &    7.53      &	 10.79       &	        &&	      &		       &	    &		 &	   &		   &		     &      \\
800    & 15.08   &  L$_{BB}$ (c)  &  	       &	   &    2.62      &	 6.36	     &	        &&	      &		       &	    &		 &	   &		   &		     &      \\
       &         &  N$_{gr}$ (d)  &  	       &	   &    173.76    &	 421.51      &	        &&	      &		       &	    &		 &	   &		   &		     &      \\
       &         &  M$_{HD}$ (e)  &  	       &	   &    103.33    &	 250.65      &	        &&	      &		       &	    &		 &	   &		   &		     &      \\
\hline
       &         &  Cont.(\%) 	  &  1.8       &  0	   & 	0	  &    0	     &  2.2	&&   0        &  0	       &  0	    &	0	 &  0	   &   0	   & 0  	     &   0	   \\ 
       &         &  F$_{BB}$ (b)  &  10.51     &	   & 		  &		     & 12.17	&&	      & 	       &	    &		 &	   &		   &		     &  	   \\
900    & 25.93   &  L$_{BB}$ (c)  &  4.54      &	   & 		  &		     & 19.46	&&	      & 	       &	    &		 &	   &		   &		     &  	   \\
       &         &  N$_{gr}$ (d)  &  175.21    &	   & 		  &		     & 750.60	&&	      & 	       &	    &		 &	   &		   &		     &  	   \\
       &         &  M$_{HD}$ (e)  &  104.19    &	   & 		  &		     & 446.34	&&	      & 	       &	    &		 &	   &		   &		     &  	   \\
\hline
       &         &  Cont.(\%) 	  &  0         &  0	   & 	0	  &    0	     &  0	&&   0        &  0	       &  0	    &	0	& 2.3	   &   0	   & 0  	    &	0     \\ 
       &         &  F$_{BB}$ (b)  &	       &	   & 		  &		     &  	&&	      & 	       &	    &		& 7.12     &		   &		    &	     \\
1200   & 97.39   &  L$_{BB}$ (c)  &	       &	   & 		  &		     &  	&&	      & 	       &	    &		& 0.07     &		   &		    &	     \\
       &         &  N$_{gr}$ (d)  &	       &	   & 		  &		     &  	&&	      & 	       &	    &		& 0.76     &		   &		    &	     \\
       &         &  M$_{HD}$ (e)  &	       &	   & 		  &		     &  	&&	      & 	       &	    &		& 0.45     &		   &		    &	     \\
\hline
       &         &  Cont.(\%) 	  &  0         &   0       &    0	  &    0	     &  5.7	&&   0        &  0	       &  0	    &	0	&  2.0     &   0	   & 0  	     &    0    \\ 
       &         &  F$_{BB}$ (b)  &	       &           &		  &		     &  2.54	&&	      & 	       &	    &		&  4.05    &		   &		     &         \\ 
1300   &  140.75 &  L$_{BB}$ (c)  &	       &           &		  &		     &  4.07	&&	      & 	       &	    &		&  0.04    &		   &		     &         \\ 
       &         &  N$_{gr}$ (d)  &	       &           &		  &		     &  28.91	&&	      & 	       &	    &		&  0.30    &		   &		     &         \\ 
       &         &  M$_{HD}$ (e)  &	       &           &		  &		     &  17.19	&&	      & 	       &	    &		&  0.18    &		   &		     &         \\ 
\hline
       &         &  Cont.(\%) 	  &  3.6       &  0.9	   & 	0	  &    0	     &  0	&&   8.2      &  3.9	       &  22.6      &	2.1	& 2.5	   &   4.5	   & 3.0	     &  13.7	   \\ 
       &         &  F$_{BB}$ (b)  &  1.21      &  1.07     &		  &		     &  	&&   5.17     & 3.70	       &  6.54      &  4.35	& 3.61     &   0.56	   & 1.13	     &  6.29	  \\ 
1400   & 197.92  &  L$_{BB}$ (c)  &  0.52      &  0.12     &		  &		     &  	&&   4.76     & 1.10	       &  39.64     &  0.12	& 0.04     &   1.32	   & 1.26	     &  8.39	  \\ 
       &         &  N$_{gr}$ (d)  &  2.64      &  0.63     &		  &		     &  	&&   24.04    & 5.54	       &  200.31    &  0.63	& 0.19     &   6.66	   & 6.36	     &  42.42	  \\ 
       &         &  M$_{HD}$ (e)  &  1.57      &  0.37     &		  &		     &  	&&   14.29    & 3.29	       &  119.11    &  0.37	& 0.11     &   3.96	   & 3.78	     &  25.22	  \\ 
\hline 
\multicolumn{2}{c}{$\sum$M$_{HD}$ (e)}& & 106  &  0.37     &  103         &  251	     & 463	&&   14       &  3	       &  119       &  0.4	& 0.75	   &    4	   & 4	             & 25	\\
\multicolumn{2}{c}{cz (f)}            & & 4507 &   2335    &      4046    &	 4046	     & 8671	&&   6579     &  3731	       &  16878     &  1157	& 700	   &  10552	   &  700	     & 7922 \\
\noalign{\smallskip}
\hline
\end{tabular}
\begin{list}{Table notes:}
\item (a) $\rm 1\times10^{-19} erg\,s^{-1}\,Hz^{-1}$; (b) $\rm 1\times10^{-27} erg\, cm^{-2}\,s^{-1}\,Hz^{-1}$; 
(c) $\rm 1\times10^{-27} erg\,s^{-1}\,Hz^{-1}$; (d) $\rm 1\times10^{43}$; (e) $\rm 1\times10^{-5} M\odot$; 
(f) From NED.  Note that as we have not detected contributions for the 
T=700\,K, 2.64 and T=1100\,K $BB$ components we left them out from the table.
\end{list}
\end{small}
\end{table*}

To compare the mass values derived for our sample and those reported
by other authors, Tab.~\ref{masses} lists the masses available in literature, determined
following \citep{bar87}. It is important to note that only three objects,
Mrk\,1239, Mrk\,766 and NGC\,7582, of the nine listed in Tab.~\ref{masses} have  
the hot dust mass estimated by means of spectroscopy. In the remaining six, the
masses were determined using photometry. The spectroscopic approach allows 
a careful subtraction of the power-law contribution and the stellar population. 
Moreover, our results have the advantage, over previous determinations, of considering 
the SP, the $FC$ and the $BB$ in the same fitting process. 
Thus, our study has increased significantly the number 
of AGNs with the mass of hot dust estimated.

\begin{table}
\renewcommand{\tabcolsep}{0.70mm}
\centering
\caption{Hot dust masses found in AGNs.}
\vspace{0.3cm}
\begin{tabular}{l c c l}
\hline
 Galaxy                  & $M_{\rm HD}$ (M$_\odot$)& Spectra & Reference \\
\hline
NGC\,7582 &              $2.8\times10^{-3}$     & yes    & \citet{rogemar08b}		      \\
Mrk\,1239 &              $2.7\times10^{-2}$     & yes    &  \citet{rom06}  	 \\
Mrk\,766   &             $2.1\times10^{-3}$     & yes    & \citet{ara05}	       \\
NGC\,1068 &              $1.1\times10^{-3}$     & no     &  \citet{marcoall00}			\\
NGC\,7469 &              $5.2\times10^{-2}$     & no     & \citet{marcoall98}			\\
NGC\,4593  &             $5.0\times10^{-4}$     & no     & \citet{santos95}		      \\ 
NGC\,3783  &             $2.5\times10^{-3}$     & no     & \citet{glass92} 		     \\
NGC\,1566  &             $7.0\times10^{-4}$     & no     &  \citet{baribaud92}		     \\
Fairall\,9 &             $2.0\times10^{-2}$     & no     & \citet{clavel89}		     \\
\hline
\end{tabular}
\label{masses}
\end{table}

 Given that the  radius of the integrated region covered by most of our data
is less than $\sim$500\,pc\footnote{Except for Arp\,102\,B, Mrk\,124 
and Mrk\,291.}, this value sets an upper limit to the size of the
hot dust emission region. We may further constrain this emission region if we 
consider that in half of our sample the spatial
resolution is $\lesssim$\,400\,pc. It means that the bulk of the hot dust is more
likely concentrated close to the central source. 

The origin of the hot dust can be the putative torus required by the unified model for AGNs 
\citep{antonucci85,antonucci93}, which is a natural dust reservoir.  This hypothesis
is further supported by the  detection of hot dust in the inner  $\sim$25\,pc of NGC\,7582 \citep{rogemar08b} and  by \citet{jaffe04}, who studied the Mid-infrared spectrum of NGC\,1068, using 
interferometry with the {\it Very Large Telescope Interferometer}-VLTI. The latter work shows that 
the 10\,\mc\ emission is due to hot dust at 800\,K concentrated in the central pc of this object, surrounded by cooler 
dust (T=300\,K) in scales of $\sim$2-4\,pc.  However, the detailed discussion of 
dust distribution around the central region of AGNs is beyond the scope of this paper and is left 
for a forthcoming work (Riffel et al. 2009, {\it in preparation}).

\section{Final Remarks}\label{finalremarks}

In this work we investigate the NIR spectra of 24 Seyfert galaxies (9 Sy\,1 and 15 Sy\,2) observed
with the IRTF SpeX, obtained in the short cross-dispersed mode. Our main focus
was the stellar population, AGN featureless continuum and dust contribution
properties, along the full wavelength coverage (0.8\mc\ - 2.4\mc). We have analysed the 
absorption features located in the NIR.
The approach followed here is based on the {\sc starlight} code, which considers the whole 
observed spectrum, continuum
and absorption features. This is the first instance where {\sc starlight} is applied to this
wavelength range. Besides, in this work we also consider for the first time hot dust as an
additional element base.

The main results can be summarised as follows. We found evidence of correlation among
the \w\ of  Si\,I\,1.59\mc\ $\times$ Mg\,I\,1.58\mc, equally for both kinds of activity. Part of the
$W\rm_{Na\,I\,2.21\mu m}$ and $W\rm_{CO\,2.3\mu m}$ strengths and the correlation between 
$W\rm_{Na\,I\,2.21\mu m}$ and $W\rm_{Mg\,I\,1.58\mu m}$ appears to be accounted for 
by galaxy inclination. For the 7 objects in
common with previous optical studies (based on the same method of analyses), 
the NIR stellar population synthesis does not reproduce well
the optical results. Our synthesis shows significant differences between Sy\,1 and Sy\,2
galaxies. The hot dust component is required to fit the $K$-band spectra of $\sim$80\%
of the Sy\,1 galaxies, and only of $\sim$40\% of the Sy\,2. Besides, about 50\% of the Sy\,2 galaxies
require a featureless component contribution in excess of 20\%, while this fraction increases
to about 60\% in the Sy\,1. Also, in about 50\% of the Sy\,2, the combined FC and $X_Y$ components
contribute with more than 20\%, while this occurs in 90\% of the Sy\,1. This suggests recent
star formation (CF05) in the central region of our galaxy sample. 
We found that the light at 1.223\mc\ in central regions of the galaxies studied here
contain a substantial fraction of intermediate-age SPs with a mean metallicity near solar. 
Moreover, our analysis confirms that the 1.1\mc\ CN band can be taken 
as an unambiguous tracer of intermediate-age stellar populations.

One consequence of this work - especially because of the simultaneous fitting of SP, $FC$ and hot 
dust components, allowing a proper analysis of each one of them - is a 150\% 
(400\% if only spectroscopic studies are considered) increase in 
the number of AGNs with hot dust detected and the mass estimated.

What emerges from this work is that the NIR may be taken as an excellent window to study the stellar
population of Sy\,1 galaxies, as opposed to the usually heavily attenuated optical range. Our
approach opens a new way to investigate and quantify the contribution of the three most important
NIR continuum components observed in AGNs.

\section*{Acknowledgements}
 We thank the anonymous referee for useful comments.
R. R. thanks to the Brazilian funding agency CAPES. ARA acknowledges 
support of the Brazilian Funding Agency CNPq under grant
311476/2006-6. The {\sc starlight} project 
is supported by the Brazilian agencies CNPq, CAPES and
FAPESP and by the France-Brazil CAPES/Cofecub program.This research has made use of
the NASA/IPAC Extragalactic Database (NED) which is operated by the
Jet Propulsion Laboratory, California Institute of Technology, under
contract with the National Aeronautics and Space Administration. We acknowledge the 
usage of the HyperLeda database (http://leda.univ-lyon1.fr).

\end{document}